\documentclass[12pt,preprint]{aastex}

\usepackage{amssymb}
\usepackage{graphicx}
\usepackage{subfigure}

\shorttitle{M31 globular cluster characterization and model comparison}
\shortauthors{Z. Fan et al.}

\begin{document}

\title{Lick Indices and Spectral Energy Distribution Analysis based on
  an M31 Star Cluster Sample: Comparisons of Methods and Models}

\author{Zhou Fan\altaffilmark{1}, Richard de Grijs\altaffilmark{2,3,4},
  Bingqiu Chen\altaffilmark{3}, Linhua Jiang\altaffilmark{2}, Fuyan
  Bian\altaffilmark{5,6}, \and Zhongmu Li\altaffilmark{7}}

\altaffiltext{1}{Key Laboratory for Optical Astronomy, National
  Astronomical Observatories, Chinese Academy of Sciences, 20A Datun
  Road, Chaoyang District, Beijing 100012, China}

\altaffiltext{2}{Kavli Institute for Astronomy and Astrophysics,
  Peking University, 5 Yi He Yuan Lu, Hai Dian District, Beijing
  100871, China}

\altaffiltext{3}{Department of Astronomy, Peking University, 5 Yi He
  Yuan Lu, Hai Dian District, Beijing 100871, China}

\altaffiltext{4}{International Space Science Institute--Beijing, 1
  Nanertiao, Zhongguancun, Hai Dian District, Beijing 100190, China}

\altaffiltext{5}{Research School of Astronomy \& Astrophysics, Mount Stromlo
  Observatory, Cotter Road, Weston ACT 2611, Australia}

\altaffiltext{6}{Stromlo Fellow}

\altaffiltext{7}{Institute for Astronomy and the History of Science
  and Technology, Dali University, Dali 671003, China}

\email{zfan@bao.ac.cn, grijs@pku.edu.cn}

\begin{abstract}
Application of fitting techniques to obtain physical parameters---such
as ages, metallicities, and $\alpha$-element to iron ratios---of
stellar populations is an important approach to understand the nature
of both galaxies and globular clusters (GCs). In fact, fitting methods
based on different underlying models may yield different results, and
with varying precision. In this paper, we have selected 22 confirmed
M31 GCs for which we do not have access to previously known
spectroscopic metallicities. Most are located at approximately one
degree (in projection) from the galactic center. We performed
spectroscopic observations with the 6.5 m MMT telescope, equipped with
its Red Channel Spectrograph. Lick/IDS absorption-line indices, radial
velocities, ages, and metallicities were derived based on the $\rm
EZ\_Ages$ stellar population parameter calculator. We also applied
full spectral fitting with the {\sc ULySS} code to constrain the
parameters of our sample star clusters. In addition, we performed
$\chi^2_{\rm min}$ fitting of the clusters' Lick/IDS indices with
different models, including the Bruzual \& Charlot models (adopting
Chabrier or Salpeter stellar initial mass functions and 1994 or 2000
Padova stellar evolutionary tracks), the {\sc galev}, and the Thomas
et al. models. For comparison, we collected their $UVBRIJK$ photometry
from the Revised Bologna Catalogue (v.5) to obtain and fit the GCs'
spectral energy distributions (SEDs). Finally, we performed fits using
a combination of Lick/IDS indices and SEDs. The latter results are
more reliable and the associated error bars become significantly
smaller than those resulting from either our Lick/IDS indices-only or
our SED-only fits.
\end{abstract}


\section{Introduction}
\label{intro.sec}

Globular clusters (GCs) are good tracers to aid in our understanding
of the formation, evolution, and interactions of galaxies. Since most
GCs were formed during the early stages of their host galaxies' life
cycles, they are often considered the fossils of the galaxy formation
and evolution processes \citep{bh00}. Since GCs are dense and
luminous, so that they can be detected at great distances, they could
be useful probes for studying the properties of distant extragalactic
systems. Since halo globular clusters (HGCs) are located at great
distances from their host galaxies' centers, they are useful to study
the dark matter distributions in their host galaxies. Another
advantage of observing HGCs is the reduced galaxy background
contribution, enabling us to achieve sufficiently high signal-to-noise
ratios (SNR) fairly easily.

The Pan-Andromeda Archaeological Survey \citep[PAndAS;][]{mcc09} has
obtained observations of the nearby galaxies Messier 31 (M31, the
Andromeda galaxy) and its companion galaxy, Messier 33 (M33), with the
MegaPrime/MegaCam camera at the 3.6 m Canada-–France-–Hawai'i
Telescope (CFHT). The survey reaches depths of $g=26.5$ mag and
$i=25.5$ mag, which has enabled discoveries of a great number of
substructures and giant stellar streams in the M31/M33 halo. These
structures are faint and spatially extended, thus making them
difficult to observe using either imaging or spectroscopy. However,
the system's HGCs may be good tracers of these structures (if they are
at least spatially related to the stellar streams), since the HGCs are
bright and characterized by centrally concentrated luminosity
profiles. They can hence be used to study the interaction between M31
and M33.

The number of M31 GCs is estimated to range from $460\pm70$
\citep{bh01} to $\sim$530 \citep{per10}, including dozens of
HGCs. \citet{h04} discovered nine previously unknown M31 HGCs based on
the Isaac Newton Telescope (INT) survey. Subsequently, \citet{h05}
discovered three new extended GCs in the halo of M31, whose nature
appears to straddle the parameter space between typical GCs and dwarf
galaxies. \citet{mac06} reported four extended, low surface brightness
star clusters in the halo of M31 based on {\sl Hubble Space Telescope}
({\sl HST})/Advanced Camera for Surveys (ACS) imaging. \citet{h07}
found 40 new, extended GCs in the halo of M31 at $\sim100$ kpc from
the galactic center, based on the INT and CFHT imaging
surveys. Recently, \citet{fan11,fan12} observed dozens of confirmed
M31 HGCs using the OptoMechanics Research, Inc. (OMR) spectrograph on
the 2.16 m telescope \citep[see,][]{fan16} at Xinglong Observatory
(National Astronomical Observatories, Chinese Academy of Sciences) in
the Fall seasons of 2010 and 2011. They estimated the ages,
metallicities, and $\alpha$-element abundances using simple stellar
population (SSP) models, as well as radial velocities, $V_r$. They
found that most HGCs are old and metal-poor. Evidence of a metallicity
gradient was also uncovered, although at a low level of significance.

Stellar population fitting techniques are important tools to constrain
the physical parameters---e.g., the ages, metallicities, and
$\alpha$-element abundances---of GCs; $\chi^2_{\rm min}$ fitting of GC
spectral energy distributions (SEDs) is an efficient way to derive
these parameters on the basis of photometry \citep[see,
  e.g.,][]{fan06,fan10,fd12,fd14,ma07,ma09,ma10,ma11,wang10}, while
spectroscopic fitting is an alternative, equally efficient approach to
deriving these parameters. In fact, many different spectroscopic
fitting techniques have been developed. One of these involves full
spectral fitting, e.g., using {\sc ULySS}
\citep{kol09,cbq}. Alternatively, one could pursue $\chi^2_{\rm min}$
fitting of various Lick/IDS indices
\citep[e.g.,][]{fan11,fan12,cbq}. Each method is associated with its
own advantages and disadvantages. Of course, the results and precision
are model-dependent. Availability of more and more useful information
obviously leads to higher precision, all else being equal. Therefore,
$\chi^2_{\rm min}$ fitting combined with SED and Lick-index fitting is
expected to provide more reliable and higher-precision results than
any of these approaches on their own.

In this paper, we compare the results of different fitting methods to
the observables of 22 M31 HGCs using SED-only and Lick-index-only
data, and their combination. This paper is organized as follows. In
Section \ref{sam.sec} we describe our sample of M31 GCs and their
spatial distribution. In Section \ref{obs.sec} we provide an overview
of our spectroscopic observations with the 6.5 m MMT telescope and the
data reduction, including our measurements of the radial velocities
and Lick indices. Subsequently, in Section \ref{fit.sec}, we derive
and discuss the ages, metallicities, and $\alpha$-element abundances
based on $\chi^2_{\rm min}$ fitting, using various models and
methods. Finally, we summarize our work and offer our conclusions in
Section \ref{sum.sec}.

\section{GC Sample Selection}
\label{sam.sec}

Our sample star clusters were mainly selected from \citet{pea10}, who
provide a catalog of 416 old, 156 young, and 373 candidate
clusters. This catalog is based on $ugriz$ and $K$-band photometry
using observations with the Wide Field Camera (WFCAM) mounted on the
United Kingdom Infrared Telescope (UKIRT) and the Sloan Digital Sky
Survey ({\sc sdss}). We selected those confirmed star clusters from
\citet{pea10} without previously determined spectroscopic
metallicities. \citet{fan08} published a spectroscopic metallicity
catalog (`SMCat') based on spectroscopic metallicities found in the
literature, specifically in \citet{per02}, \citet{hbk91}, and
\citet{bh00}. Their catalog contains 295 entries. In addition, we also
include the complementary spectroscopic metallicities of
\citet{gall09}, leading to a final database of 329 metallicities.

We next excluded star clusters with previous spectroscopic metallicity
determinations. Thus, we obtained a catalog of 102 confirmed GCs which
lack spectroscopic metallicities or radial velocities. In addition,
since we are interested in halo star clusters and to achieve
sufficiently high SNRs for the observations, we removed star clusters
located inside a projected distance of $r_{\rm p}=1$ degree from the
the galaxy's center. Since the local background near the galaxy center
is too luminous for our observations, we were left with only 35
GCs. Of these, we observed 17 randomly chosen objects given the
limited observing time we had access to, thus minimizing selection
effects. The magnitudes of our sample star clusters range from $V =
15.5$ mag to $V = 18.2$ mag. For comparison, we also included five
star clusters (SK001A, B423, B298, B006, and B017) which have their
radial velocities listed in the Revised Bologna Catalogue of M31 GCs
and candidates \citep[RBC v.4,][]{gall04,gall06,gall09}. B423 is
listed as a confirmed star cluster in the RBC v.4 but classified as a
star cluster candidate in \citet{pea10}. It does not have any
previously obtained spectroscopic information.

The observational information of our sample GCs is listed in
Table~\ref{t1.tab}, which includes the cluster identification
(Col. 1), coordinates (R.A. and Dec. in Cols. 2 and 3, respectively),
projected radii from the galaxy center, $r_{\rm p}$, in kpc (Col. 4),
$V$-band magnitude (Col. 5), exposure time (Col. 6), and observation
date(s) (Col. 7). In fact, the coordinates, $r_{\rm p}$, and $V$-band
magnitudes were taken from the RBC v.5. The projected radii were
calculated for the M31 center coordinates \citep[J2000: 00:42:44.31,
  $+$41:16:09.4;][]{per02}, a position angle (PA) of $38^{\circ}$, and
a distance $d=785$ kpc \citep{mcc05}. SK001A is classified as a
confirmed star cluster in the RBC v.4, but in the updated RBC v.5 it
has been re-classified as a star. Since we performed our observations
before the RBC v.5 was updated, SK001A was included in our observing
campaign.

\begin{figure}
\resizebox{\hsize}{!}{\rotatebox{0}{\includegraphics{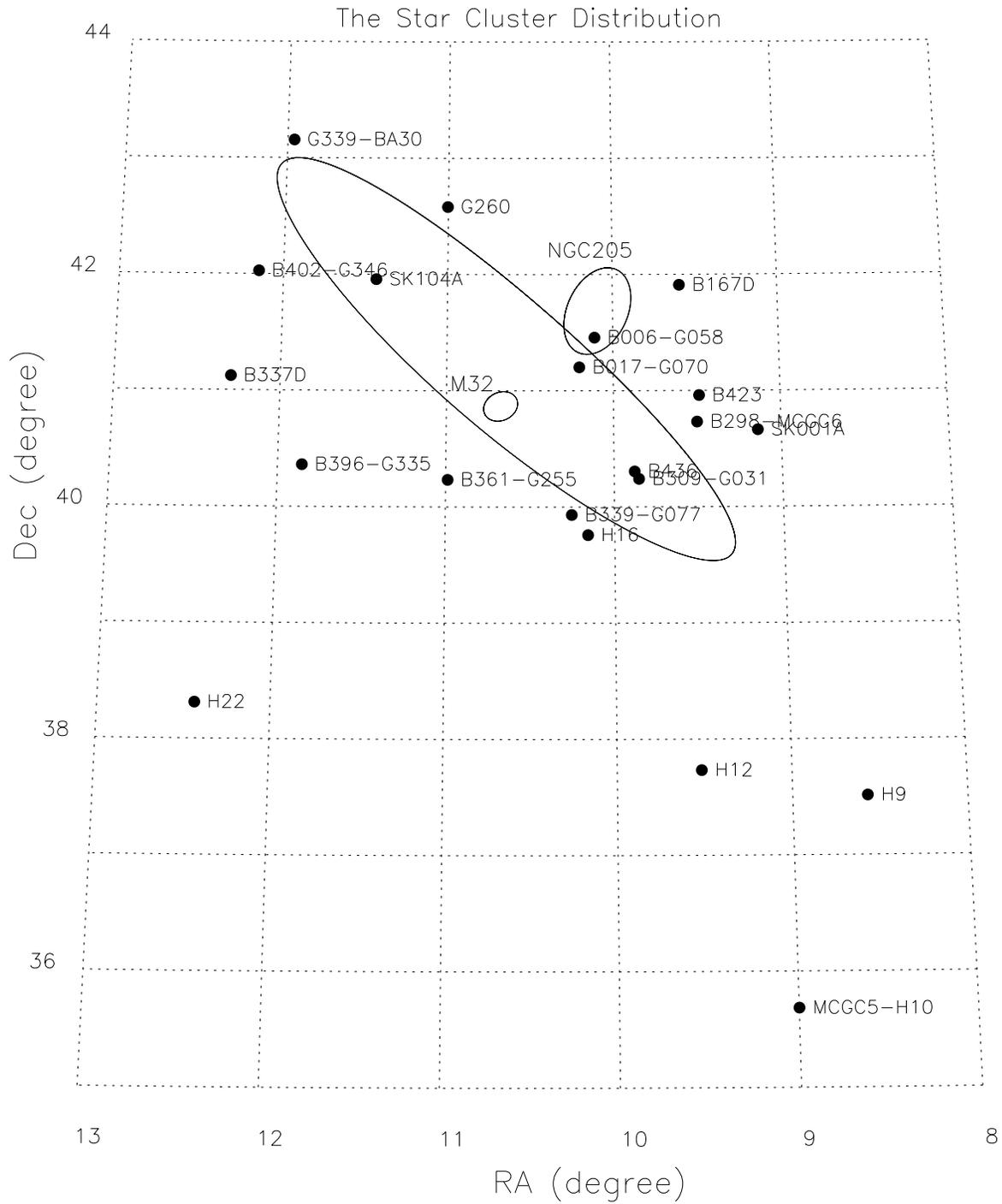}}}
\caption{Spatial distribution of the M31 GCs. All our sample GCs are
  shown as solid bullets marked with their names. The large ellipse is
  the M31 disk/halo boundary as defined in \citet{rac91}. M33 and NGC
  205 are also marked.}
\label{fig1}
\end{figure}

In Fig.~\ref{fig1} we show the spatial distribution of our sample star
clusters. The large ellipse is the M31 disk/halo boundary as defined
by \citet{rac91}; the two smaller ellipses trace out NGC 205 and
M32. Note that most of our sample GCs are located in the projected
direction of the M31 halo, which could help us to better understand
the nature of the galaxy's halo with the enlarged cluster sample of
\citet{fan11,fan12}.

\section{Observations and Data Reduction}
\label{obs.sec}

Spectroscopic observations were carried out with the 6.5 m MMT/Red
Channel Spectrograph from 2010 October 31 to 2010 November 2 and on
2011 November 4. The telescope is located on Mt. Hopkins in Arizona
(USA) at an altitude of 2581 m. The exposure times used ranged from
480 s to 1800 s, depending on the cluster brightness. The median
seeing was $\sim0''.98$ and we adopted a slit aperture of
$0''.75\times180''$. The CCD's size is 450$\times$1032 pixels$^2$. It
is characterized by a gain of 1.3 e$^{-}$ ADU$^{-1}$, with a readout
noise (RN) of 3.5 e$^{-}$. A grating with 600 $l$ mm$^{-1}$ with a
blaze 1st/4800 was used. The spectral resolution was $R=960$ for slit
of 1'' and a central wavelength of 4701{\AA}; the dispersion was
1.63{\AA} pixel$^{-1}$. 

The spectroscopic data were reduced following standard procedures with
the NOAO Image Reduction and Analysis Facility ({\sc iraf} v.2.15)
software package. First, all spectral images were checked
carefully. Next, we performed bias combination with {\tt zerocombine}
and bias correction with {\tt ccdproc}. Subsequently, flat-field
combination, normalization, and corrections were done using {\tt
  flatcombine}, {\tt response}, and {\tt ccdproc},
respectively. Cosmic rays were removed using {\tt cosmicrays}. We
extracted both the star cluster spectra and those obtained from a
comparison arc lamp with {\tt apall}. Wavelength calibration was done
with helium/argon-lamp spectra taken at both the beginning and the end
of observations during each night. The spectral features of the
comparison lamps were identified with {\tt identify} and wavelength
calibration was done with {\tt refspectra}. We next used {\tt dispcor}
for dispersion correction and to resample the spectra. Flux
calibration was performed based on observations of four Kitt Peak
National Observatory (KPNO) spectral standard stars \citep{mass88}. We
applied {\tt standard} and {\tt sensfunc} to combine the standards and
to determine the sensitivity and extinction. Atmospheric extinction
was corrected for using the mean extinction coefficients pertaining to
KPNO. Finally, we applied {\tt calibrate} to correct for the
extinction and complete the flux calibration.

\begin{figure}
  \resizebox{\hsize}{!}{\rotatebox{0}{\includegraphics{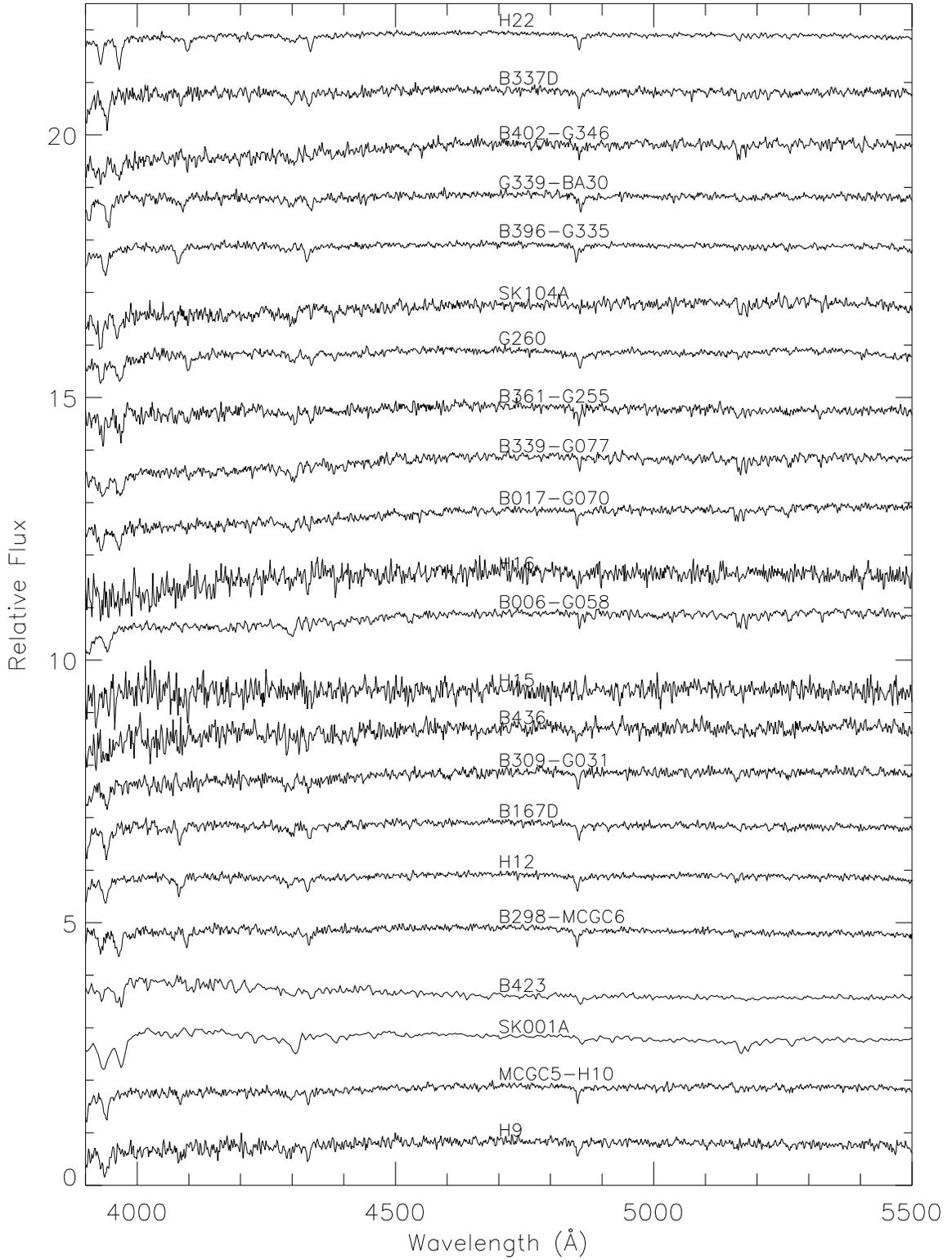}}}
  \caption{Normalized spectra of our sample GCs taken with the MMT's Red
    Channel Spectrograph.}
  \label{fig1a}
\end{figure}

Figure~\ref{fig1a} shows the normalized, calibrated spectra of our 
sample GCs, identified by their names. The signal-to-noise ratios of most 
GCs are high, except for those of a few faint clusters such as H16, H15, and
B436.

\section{Fitting, Analysis, and Results}
\label{fit.sec}

\subsection{Full-Spectrum Fitting with ULySS}
\label{fsp.sec}

{\sc ULySS} \citep{kol09} was used for the full spectral fitting of
the ages and metallicities. The \citet{vaz} SSP models adopted are
based on the {\sc miles} (Medium-resolution INT Library of Empirical
Spectra) spectral library \citep{sb06}. The wavelength coverage ranges
from 3540.5{\AA} to 7409.6{\AA} at a nominal resolution full width at
half maximum, FWHM = 2.3{\AA}. The solar-scaled theoretical isochrones
of \citet{gi00} were adopted and we used the stellar initial mass
function (IMF) of \citet{sal} for the fitting. The age coverage was
$10^8$--$1.5\times10^{10}$ yr and the metallicity ranged from [Fe/H] =
$-2.32$ dex ($Z=0.0004$) to [Fe/H] = $+0.22$ dex ($Z=0.03$). An
independent SSP model set, {\sc pegase-hr}, is provided by
\citet{leb}, which is based on the empirical spectral library {\sc
  elodie} \citep[e.g.,][]{ps01,pr07}. The {\sc elodie} wavelength
coverage spans the range from 3900{\AA} to 6800{\AA}. The spectral
resolution is $R\sim10,000$ with a FWHM of $\sim0.55${\AA}. The
effective temperature, $T_{\rm eff}$, range is 3100--50,000 K, the
gravity, $log~g$, ranges from $-0.25$ dex to 4.9 dex, and the
metallicity $\rm [Fe/H]$ ranges from $-3$ dex to $+1$ dex. The flux
calibration accuracy is 0.5--2.5\% from narrow to broad bands. We
adopted the {\sc pegase-hr} SSP models with a \citet{sal} IMF. The
model ages we adopted cover the range $10^7$--$1.5\times10^{10}$ yr,
and the metallicity ranges from [Fe/H] = $-2.0$ dex ($Z=0.0004$) to
[Fe/H] = $+0.4$ dex ($Z=0.05$).

\begin{figure}
  \resizebox{\hsize}{!}{\rotatebox{0}{\includegraphics{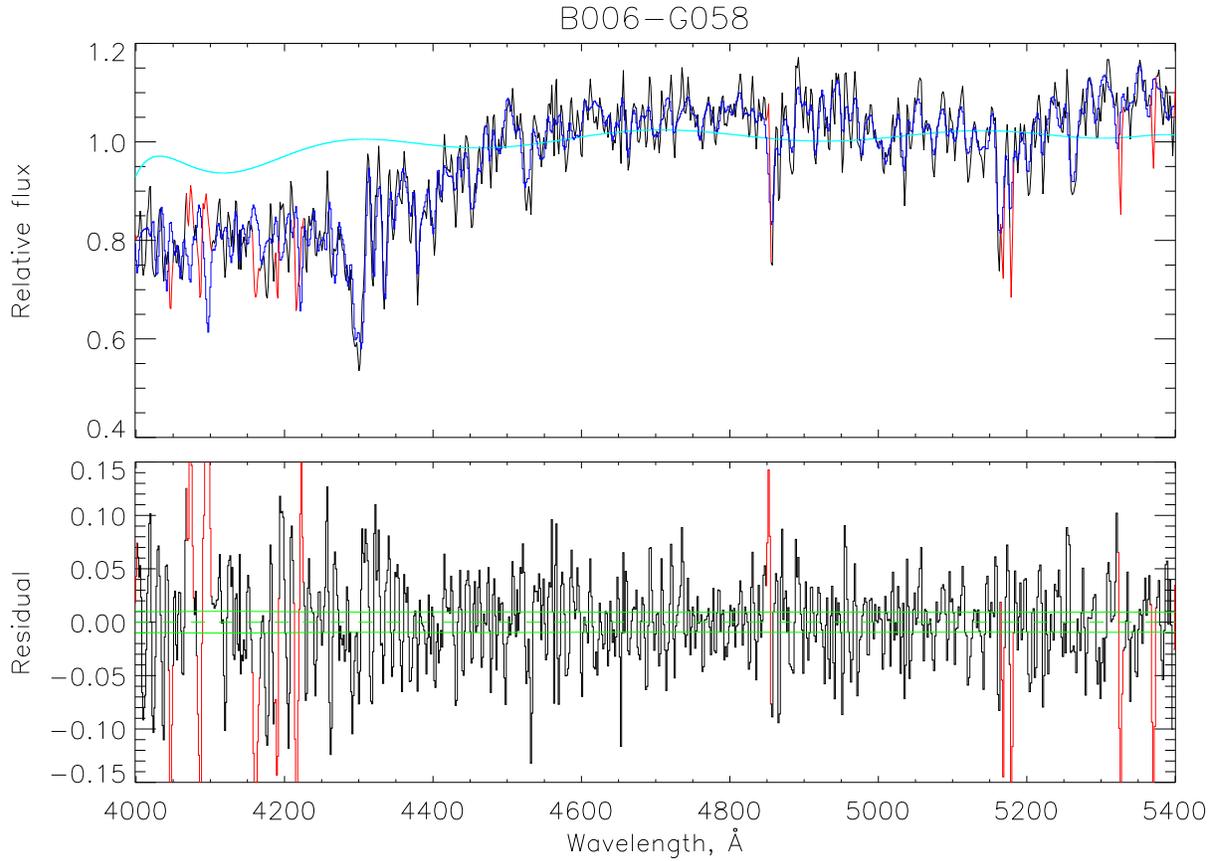}}}
  \caption{Example {\sc ULySS} full-spectrum fit for the cluster
    B006--G058. The top panel shows the observed spectrum in black;
    the best-fitting spectra fromcthe {\sc pegase-hr} model is shown
    in blue; the outliers of the fits are reproduced in red. The cyan
    lines delineate the multiplicative polynomials. The bottom panels
    are the fractional residuals of the best fits, where the dashed
    and solid lines in green denote, respectively, zero and the
    1$\sigma$ deviations, with the latter calculated from the
    (in)variance of the input (observed) spectra. }
  \label{fig2}
\end{figure}

Figure~\ref{fig2} shows the observed MMT spectroscopy of the star
cluster B006--G058 combined with the best-fitting model from {\sc
  pegase-hr}. For the full spectral fitting we only consider the
wavelength range $\sim4000$--$5400${\AA} for SNR reasons. In the top
panel, the observed spectrum is shown in black, the best-fitting
spectrum from the {\sc pegase-hr} models is rendered in blue, and the
outliers are shown in red. The cyan lines delineate the multiplicative
polynomials. The bottom panels are the fractional residuals of the
best fits. The dashed and solid lines in green denote zero and the
1$\sigma$ deviations, respectively, which were calculated based on the
variance of the input (observed) spectrum. For all sample star
clusters, the results of the fits are good and the residuals are
small.

The radial velocities and their associated uncertainties are fitted
with {\sc ULySS} based on both the \citet{vaz} and {\sc pegase-hr} SSP
models. The results are included in Table~\ref{t2.tab}. For
comparison, we also list the $V_r$ values from the RBC v.5. For most
GCs, the values resulting from the \citet{vaz} and {\sc pegase-hr} SSP
models are rather similar. The median $V_r$ values are $-353$ km
s$^{-1}$ for the \citet{vaz} models and $-351$ km s$^{-1}$ for the
{\sc pegase-hr} SSP models. Both yield a velocity difference of $\sim
-50$ km s$^{-1}$ relative to the systematic velocity of M31, $\sim
-300$ km s$^{-1}$.  Note that for most clusters, the $V_r$ values
derived from our {\sc ULySS} fits are consistent with those from the
RBC v.5, except for a few clusters (G339--BA30, G260, and SK001A). We
carefully checked the spectroscopy and concluded that the fit results
are reasonable and stable. The RBC v.5 data were collected from
various sources in the literature, some of which may have large
measurement uncertainties but which are not listed in the RBC v.5. The
systematic differences between the observed velocities and the catalog
velocity are $-63\pm137$ km s$^{-1}$ for the \citet{vaz} models and
$-61\pm147$ km s$^{-1}$ for the {\sc pegase-hr} models. This indicates
that our measurements agree well with those in the RBC v.5, since the
systematic differences between our measurements and the published
values are not significant. For some star clusters, the differences
between our measurements and the published values are relatively
large, which may be due to the different kinds or numbers of spectral
features or methods used for the measurements and analysis.

Table~\ref{t3.tab} summarizes the resulting logarithmic ages and the
metallicities derived based on the Lick absorption-line indices and
fitted with {\sc ULySS}, for both the \citet{vaz} and {\sc pegase-hr}
SSP models. As expected, the ages and metallicities resulting from
both model sets are essentially the same. All our sample GCs are old
and metal-poor: for the \citet{vaz} SSP models, $\rm age_{mean}=11.7$
Gyr (with an r.m.s. spread of $3.9$ Gyr) and $\rm [Fe/H]_{mean}=-1.37$
dex (r.m.s. = $0.60$ dex); for the {\sc pegase-hr} models, $\rm
age_{mean}=12.9$ Gyr (r.m.s. = $6.4$ Gyr) and $\rm
[Fe/H]_{mean}=-1.41$ dex (r.m.s. = $0.62$ dex), which is consistent
with previous results.

Figure~\ref{fig3} displays the relationship between the metallicities
and the ages of our sample GCs fitted with {\sc ULySS} (from
Table~\ref{t3.tab}). The associated error bars are also shown; they
are relatively small. The left- and right-hand panels show the
parameters resulting from the \citet{vaz} and the {\sc pegase-hr} SSP
models, respectively. The results are rather similar. All star
clusters are old and most are metal-poor, since most of the sample
clusters are located in the halo of the galaxy.

\subsection{Lick-Index Fitting with $\rm EZ\_Ages$}
\label{ez.sec}

$\rm EZ\_Ages$ is an automated stellar population analysis tool
written in {\sc idl}.\footnote{The Interactive Data Language ({\sc
    idl}) is licensed by Research Systems Inc., of Boulder, CO, USA.}
The code is used to compute the mean light-weighted ages,
metallicities [Fe/H], and the elemental abundances [Mg/Fe], [C/Fe],
[N/Fe], and [Ca/Fe] from integrated spectra for various unresolved
stellar populations \citep{gs08}. Recently, \citet{cbq} successfully
tested the $\rm EZ\_Ages$ package by fitting Large Sky Area
Multi-Object Fibre Spectroscopic Telescope (LAMOST) spectroscopy of
Galactic GCs with known ages and chemical compositions, which they
subsequently applied to a sample of M31 star clusters.

We applied the $\rm EZ\_Ages$ code to estimate the Lick indices as
well as the ages and abundances of our sample GCs. Before measuring
the Lick indices, the resolution was adjusted with a variable-width
Gaussian kernel following the definition of \cite{wo97}, i.e.,
11.5{\AA} at 4000{\AA}, 9.2{\AA} at 4400{\AA}, 8.4{\AA} at 4900{\AA},
8.4{\AA} at 5400{\AA}, and 9.8{\AA} at 6000{\AA}. Since the wavelength
coverage of the MMT's Red Channel Spectrograph is
$\sim3900$--$5500${\AA}, we measured all 20 Lick indices defined in
this regime. The measurements were strictly done following
\citet{w94a} and \citet{wo97}. The uncertainty in each index was
estimated following \citet[][their Eqs 11--18]{car98}.

Table~\ref{t4.tab} lists the ages and metallicities derived with $\rm
EZ\_Ages$. Note that many GCs do not have age or metallicity values
owing to fitting failures of the $\rm EZ\_Ages$ code. Since $\rm
EZ\_Ages$ does not perform model extrapolation, stellar populations
with line index measurements outside the model grid are excluded from
the analysis. This applies to models with metallicities $\rm [Fe/H]<
-1.3$ dex or $\rm [Fe/H]> +0.2$ dex. The metallicities $\rm
[Fe/H]_{[MgFe]}$ are also given in Table~\ref{t4.tab}; they were
derived from $\rm [MgFe]$; $\rm [MgFe] = \rm \sqrt{Mg{\it b} \times
  \langle Fe \rangle}$, where $\rm \langle Fe
\rangle=(Fe5270+Fe5335)/2$. Thus, the metallicity can be calculated
following \citet{gall09}, using
\begin{equation}
  \rm [Fe/H]_{[MgFe]}=-2.563+1.119[MgFe]-0.106[MgFe]^2\pm0.15 dex (r.m.s.).
  \label{eq1}
\end{equation}

Similarly, the metallicities $\rm [Fe/H]_{\langle Fe \rangle}$,
derived from $\rm \langle Fe \rangle=(Fe5270+Fe5335)/2$ following
\citet{cw11}, are also listed in Table~\ref{t4.tab}.

\subsection{Lick Index Fitting with Stellar Population Models}
\label{res.sec}

\citet{tmb} provided stellar population models which included Lick
absorption-line indices for various elemental-abundance ratios. The
model suite's age coverage ranges from 1 Gyr to 15 Gyr and the
metallicities span from 1/200 to 3.5 times solar abundance. These
models are based on the standard models of \citet{mar98} and the input
stellar evolutionary tracks are from \citet{ccc97} and
\citet{bo97}. The \citet{sal} stellar IMF was adopted. Subsequently,
\citet{tmk} improved the models by including higher-order Balmer
absorption-line indices and found that these indices are sensitive to
changes in the $\rm \alpha/Fe$ ratio for supersolar metallicities. The
updated stellar population models for Lick absorption-line indices of
\citet{tmj} represent an improvement with respect to \citet{tmb} and
\citet{tmk}; they are based on the {\sc miles} stellar library
\citep{sb06}. The model provides a higher spectral resolution
appropriate for {\sc miles} and {\sc sdss} spectroscopy with
calibrated fluxes. The models cover ages from 0.1 Gyr to 15 Gyr (0.1,
0.2, 0.4, 0.6, 0.8, 1, 2, 3, 4, 5, 6, 7, 8, 9, 10, 11, 12, 13, 14, and
15 Gyr), the metallicity ranges from $\rm [Z/H]=-2.25$ dex to 0.67 dex
($-2.25$, $-1.35$, $-0.33$, 0.0, 0.35, and 0.67 dex), and the
$\alpha$-element ratio $\rm [\alpha/Fe]$ spans from $-0.3$ dex to 0.5
dex ($-0.3$, 0.0, 0.3, and 0.5 dex). The models are based on the
evolutionary synthesis code of \citet{m05}. The stellar evolutionary
tracks adopted come from \citet{ccc97} and \citet{gi00} (i.e., the
Padova models), especially for $\rm [Z/H]=0.67$ dex. We fitted the
absorption-line indices with the models of \citet{tmj}, adopting the
stellar evolutionary tracks of \citet{ccc97}. In fact, the models were
interpolated to allow for smaller parameter intervals to improve the
resolution of the parameters. Cubic spline interpolations were
adopted, using equal step lengths, to obtain a grid of 150 ages (from
0.1 Gyr to 15 Gyr), 31 $\rm [Z/H]$ values (from $-2.25$ dex to 0.67
dex), and 31 $\rm [\alpha/Fe]$ ratios (from $-0.3$ dex to $0.5$ dex),
which results in fits that are smoother and more continuous.

As mentioned in Section~\ref{ez.sec}, because of the limited
wavelength coverage of our sample GCs, we only measured 20 Lick
absorption-line indices. Since the Lick indices are not sensitive to
reddening, as opposed to our SED fits, we did not consider the
reddening in our Lick-index fitting. Similarly to \citet{fan11,fan12},
the ages $t$, metallicities $\rm [Z/H]$, and the $\rm [\alpha/Fe]$
ratios can be determined by comparing the interpolated stellar
population models with observational spectral indices by employing the
$\chi^2$-minimization method,
\begin{equation}
  \chi^2_{\rm min}={\rm
    min}\left[\sum_{i=1}^{20}\left({\frac{L_{\lambda_i}^{\rm
	  obs}-L_{\lambda_i}^{\rm mod}(t,\rm [Z/H],[\alpha/Fe])}
      {\sigma_{L,i}}}\right)^2\right],
\label{eq2}
\end{equation}
where $L_{\lambda_i}^{\rm mod}(t,\rm [Z/H],[\alpha/Fe])$ is the
$i^{\rm th}$ Lick absorption-line index in the stellar population
model for $t$, $\rm [Z/H]$, and $\rm [\alpha/Fe]$; $L_{\lambda_i}^{\rm
  obs}$ represent the observed, calibrated Lick absorption-line
indices from our measurements and the errors estimated from our fits
are
\begin{equation}
  \sigma_{L,i}^{2}=\sigma_{{\rm obs},L,i}^{2}+\sigma_{{\rm mod},L,i}^{2}.
\label{eq3}
\end{equation}
Here, $\sigma_{{\rm obs},i}$ is the observational uncertainty;
$\sigma_{{\rm mod},i}$ is the uncertainty associated with the model
Lick indices, provided by \citet{tmj} and Table~5 of \citet{bc03}.

Table~\ref{t5.tab} includes the resulting ages, metallicities, and
$\rm [\alpha/Fe]$ ratios for the evolutionary tracks of \citet{ccc97}
and the Padova models of \citet{tmj}. Figure~\ref{fig4} shows the
relationship between the ages and metallicities derived from the
\citet{tmj} models, which are also included in Table~\ref{t5.tab}.
Most GCs are older than or approximately 10 Gyr, and most
metallicities are lower than $\rm [Z/H]=-1$ dex. This implies that our
sample GCs are old and metal-poor, which is in agreement with the
previous results we obtained using $\rm EZ\_Ages$, included in
Table~\ref{t4.tab}. Thus, this indicates that these HGCs formed during
the early stages of galaxy formation, which agrees well with previous
conclusions.

The Galaxy Evolutionary Synthesis Models \citep[{\sc
    galev};][]{lf06,kot09} are evolutionary synthesis models that can
be used to simultaneously explore both the chemical evolution of the
gas and the spectral evolution of the stellar content in star clusters
or galaxies. The models provide photometry and Lick absorption-line
indices for different types of galaxies (i.e., different
star-formation rates) or SSPs (single burst). The models provide 5001
ages from 4 Myr to 20 Gyr, and seven metallicities ($Z=0.0001$,
0.0004, 0.001, 0.004, 0.008, 0.02, and 0.05). As done previously, we
interpolate the metallicities to a grid of 51 values which can yield
more accurate results than the basic model set. The uncertainties in
the Lick indices for the models in Eq.~(\ref{eq3}) are estimated using
Equation (3) of \citet{lf06}, while the observed uncertainties come
from the estimated $\rm EZ\_Ages$ fits to the observations. The fit
results are shown in Table~\ref{t6.tab} and the age--metallicity
distribution is shown in Fig.~\ref{fig5}. Although more than half of
the star clusters are older than 1 Gyr, there are still numbers of
star clusters that are younger than 1 Gyr but associated with large
uncertainties. This is different from our results based on
\citet{tmj}. In fact, \citet{fd12} already pointed out that the {\sc
  galev} models are mainly suitable for young stellar populations;
they usually produce younger ages than other models.

The evolutionary stellar population synthesis models of
\citet[][hereafter BC03]{bc03} do not only provide spectra and SEDs
for different physical parameters, but also Lick/IDS absorption-line
indices. The models include the 1994 and 2000 Padova stellar
evolutionary tracks as well as the \citet{sal} and \citet{chab03}
IMFs. The wavelength coverage ranges from 91{\AA} to 160 $\mu$m. For
the Padova 1994 tracks, models for six metallicities ($Z=0.0001$,
0.0004, 0.004, 0.008, 0.02, and 0.05) are provided, while for the
Padova 2000 tracks, models of six different metallicities ($Z=0.0004$,
0.001, 0.004, 0.008, 0.019, and 0.03) are given. In total, there are
221 ages from 0 to 20 Gyr in unequally spaced time steps. Here we also
interpolate the models to attain smaller intervals of the parameter
space (51 metallicities with equal steps in logarithmic space) to
obtain more accurate results. Similar to the fits based on the
\citet{tmj} models, Eqs~(\ref{eq2}) and (\ref{eq3}) were adopted. In
fact, we tried different combinations of Padova 1994/2000 stellar
evolutionary tracks and IMFs. The results are given in
Table~\ref{t7.tab} and the relevant plots are shown in
Fig.~\ref{fig6}. It is clear that for the same stellar evolutionary
tracks adoption of a different IMF does not affect the results
(including the uncertainties) significantly. Note that the
best-fitting metallicity range for the Padova 2000 tracks is not as
wide as that obtained from the Padova 1994 tracks because of the
models' limitations.

\subsection{Combination with SEDs and Comparisons with the BC03 models}
\label{sed.sec}

We also acquired the clusters' $UBVRIJHK$ photometry from the RBC v.5
and performed SED fitting to the sample GCs. Reddening affects the SED
results more significantly than for the Lick/IDS indices. The
reddening values for our sample star clusters were adopted from
\citet{fan08} where available. If they were not available, we used
$E(B-V)=0.1$ mag \citep[e.g.,][]{van69,fro80} instead, a
representative value of the Galactic foreground reddening in the
direction of M31. The extinction $A_{\lambda}$ can be computed using
the equations of \citet{ccm89}, and we adopted a typical foreground
Milky Way extinction law, $R_V = 3.1$. We fitted the SEDs using
\begin{equation}
  \chi^2_{\rm min}={\rm
    min}\left[\sum_{j=1}^{8}\left({\frac{M_{\lambda_j}^{\rm
	  obs}-M_{\lambda_j}^{\rm mod}(t,\rm [Z/H],[\alpha/Fe])}
      {\sigma_{M,j}}}\right)^2\right],
  \label{eq4}
\end{equation}
where $M_{\lambda_j}^{\rm mod}(t,\rm [Z/H],[\alpha/Fe])$ is the
$j^{\rm th}$ magnitude provided in the stellar population model for
age $t$, metallicity $\rm [Z/H]$, and $\rm [\alpha/Fe]$;
$M_{\lambda_i}^{\rm obs}$ represents the observed dereddened magnitude
in the $j^{\rm th}$ band.

Similar to Eq.~\ref{eq3}, Eq.~\ref{eq5} represents the errors
associated with our fits,
\begin{equation}
  \sigma_{M,j}^{2}=\sigma_{{\rm obs},M,j}^{2}+\sigma_{{\rm mod},M,j}^{2},
\label{eq5}
\end{equation}
where $\sigma_{M,j}$ is the magnitude uncertainty in the $j^{\rm th}$
filter. We estimated the photometric errors in the RBC v.5 magnitudes.
We adopted 0.08 mag in $U$, 0.05 mag in $BVRI$, 0.1 mag in $J$, and
0.2 mag in $HK$ \citep{gall04}. The model errors adopted were 0.05
mag, which is the typical photometric error for the BC03 and {\sc
  galev} SSP models \citep[e.g.,][]{fan06,ma07,ma09,wang10,fd14}.

The results of our SED fits are included in Table~\ref{t7.tab} and the
relevant plots are shown in Fig.~\ref{fig7}. Similarly to
Fig.~\ref{fig6}, we tried different combinations of the Padova
1994/2000 stellar evolutionary tracks and IMFs. It is clear that for
the same stellar evolutionary tracks, adoption of a different IMF does
not affect the results (including the uncertainties) significantly. We
found that the best-fitting ages are old for all clusters and the
scatter in the parameter distribution is smaller than that resulting
from the Lick absorption-line index fitting in Fig.~\ref{fig6}.

The mathematical approach to simultaneously minimize the Lick
absorption-line indices and SEDs with respect to the stellar
population models is given by
\begin{equation}
  \chi^2_{\rm min}={\rm
    min}\left[\sum_{i=1}^{20}\left({\frac{L_{\lambda_i}^{\rm
	  obs}-L_{\lambda_i}^{\rm mod}(t,\rm [Z/H],[\alpha/Fe])}
      {\sigma_{L,i}}}\right)^2+\sum_{j=1}^{8}\left({\frac{M_{\lambda_j}^{\rm\ obs}-M_{\lambda_j}^{\rm
	  mod}(t,\rm [Z/H],[\alpha/Fe])}
      {\sigma_{M,j}}}\right)^2\right],
  \label{eq6}
\end{equation}
where $L_{\lambda_i}^{\rm mod}(t,\rm [Z/H],[\alpha/Fe])$ and
$M_{\lambda_j}^{\rm mod}(t,\rm [Z/H],[\alpha/Fe])$ are the $i^{\rm
  th}$ Lick absorption-line indices and the $j^{\rm th}$ magnitudes in
the relevant stellar population model for age $t$, metallicity $\rm
[Z/H]$, and $\rm [\alpha/Fe]$, respectively; $L_{\lambda_i}^{\rm obs}$
and $M_{\lambda_j}^{\rm obs}$ represent the observed, calibrated
$i^{\rm th}$ Lick absorption-line indices and the observed, dereddened
magnitude in the $j^{\rm th}$ band, respectively. The errors
associated with the observations in our fits are given by
Eqs~(\ref{eq3}) and (\ref{eq5}).

In order to compare the results obtained from the different
approaches, we also tried different combinations of Padova 1994/2000
stellar evolutionary tracks and IMFs. The final results are included
in Table~\ref{t7.tab} and a plot of the age--metallicity distribution
is shown in Fig.~\ref{fig8}. Note that the results seem better than
those shown in Fig.~\ref{fig7}, since the uncertainties in the
parameters are significantly smaller than those resulting from the
SED-only fitting method. They also look better than those from the
Lick-indices-only fits in Fig.~\ref{fig6}. It is clear that the
uncertainties in the parameters and the scatter are much smaller. The
HGCs are expected to be metal-poor and old, which is confirmed by the
results of both our fits with {\sc ULySS} and those based on the
\citet{tmj} models. Therefore, we conclude that combined fits of SEDs
and Lick indices can significantly improve the reliability and
accuracy of the results.

Figure~\ref{fig9} compares the ages resulting from the different
models. For all panels, the $x$ axis is the age resulting from the
combined fits of the Lick indices and SEDs, adopting the BC03 models,
the Padova 1994 stellar evolutionary tracks, and a \citet{chab03} IMF,
a combination we use as our standard for comparison. Along the $y$
axis we display, at the top left, the results from the BC03 models,
the Padova 1994 tracks, and \citet{chab03} IMF fits of the Lick
indices only. Fits based on only the Lick indices agree well with the
combined fits, at least if we use the same models, the same tracks,
and the same IMF, except for a few outliers.

At the top right, we show the results for the BC03 models, the Padova
1994 tracks, and \citet{chab03}) IMF fits to the SEDs only. It is
clear that the fits based on SEDs only agree well with the results
from the combined fits for the same models, the same tracks, and the
same IMF. The agreement is even better than that in the top left-hand
panel. In the middle left-hand panel, we show the results from fits
using the BC03 models, the Padova 2000 tracks, and a \citet{chab03}
IMF for the combination of SEDs and Lick indices. The models, methods,
and IMFs used are exactly the same, except for our adoption of the
Padova 1994/2000 evolutionary tracks, indicating that any differences
between the two sets of evolutionary tracks are rather insignificant
as regards the resulting ages. In the middle right-hand panel, we show
the {\sc galev} model fits to the Lick indices only. A few of the
`old' star clusters in the BC03 models are considered `young' in the
{\sc galev} models, which is noted in Fig.~\ref{fig6} and explained in
\citet{fd12}. At the bottom left, the results from \citet{tmj} model
fits to only the Lick indices are shown and at the bottom right we
show the results from the {\sc ULySS} models combined with \citet{vaz}
full-spectrum fits. Both the results from the \citet{tmj} and those
based on the {\sc ULySS} fits using the \citet{vaz} SSP models agree
well with those from the BC03 models with the combined fits of the
Lick indices and SEDs.

Figure~\ref{fig10} is the same as Fig.~\ref{fig9}, but for the
metallicity. The $x$ axis represents the metallicities obtained from
fitting the combination of Lick indices and SEDs using the BC03
models, the Padova 1994 stellar evolutionary tracks, and a
\citet{chab03} IMF. As regards the $y$ axis, at the top left, we show
the results of the BC03 models, the Padova 1994 tracks, and
\citet{chab03} IMF fits to only the Lick indices. The agreement is
good, except for a few outliers. At the top right, results for the
BC03 models, Padova 1994 tracks, and \citet{chab03} fits to only the
SEDs are shown. The agreement is not as good as that for the fits to
only the Lick indices in the top left-hand panel. In the middle
left-hand panel, we show the results for the BC03 models (Padova 2000
models and \citet{chab03} IMF), adopting combined fits of Lick indices
and SEDs. Again, the Padova 1994 and Padova 2000 tracks agree well
with each other, except that the lower limit to the metallicity for
the Padova 2000 tracks is higher than that for the Padova 1994
tracks. In the middle right-hand panel, {\sc galev} model fits to only
the Lick indices are shown. The metallicities resulting from the two
models do not agree as well as for the other models. At the bottom
left, we show the \citet{tmj} model fits to only the Lick indices. The
\citet{tmj} results agree very well with those from the BC03 models
for the combined fits. At the bottom right, one sees the results from
the {\sc ULySS} models and the \citet{vaz} full-spectrum fits. For the
metal-poor clusters, the agreement between the two model sets is good,
but the scatter is significantly larger than for the other models.

\section{Summary and Conclusions}
\label{sum.sec}

We have selected 22 confirmed M31 GCs from \citet{pea10}, most of
which are located in the halo of the galaxy in an extended
distribution out to $\sim80$ kpc from the galaxy center. We obtained
our observations with the 6.5 m MMT/Red Channel Spectrograph; the
spectral resolution was $R=960$ for a slit width of $1''$. Since most
of these star clusters are located in the halo of the galaxy, the sky
background was dark. Thus, they could be observed with high SNRs
compared with their counterparts in the galaxy's disk.

\begin{figure}
  \resizebox{\hsize}{!}{\rotatebox{0}{\includegraphics{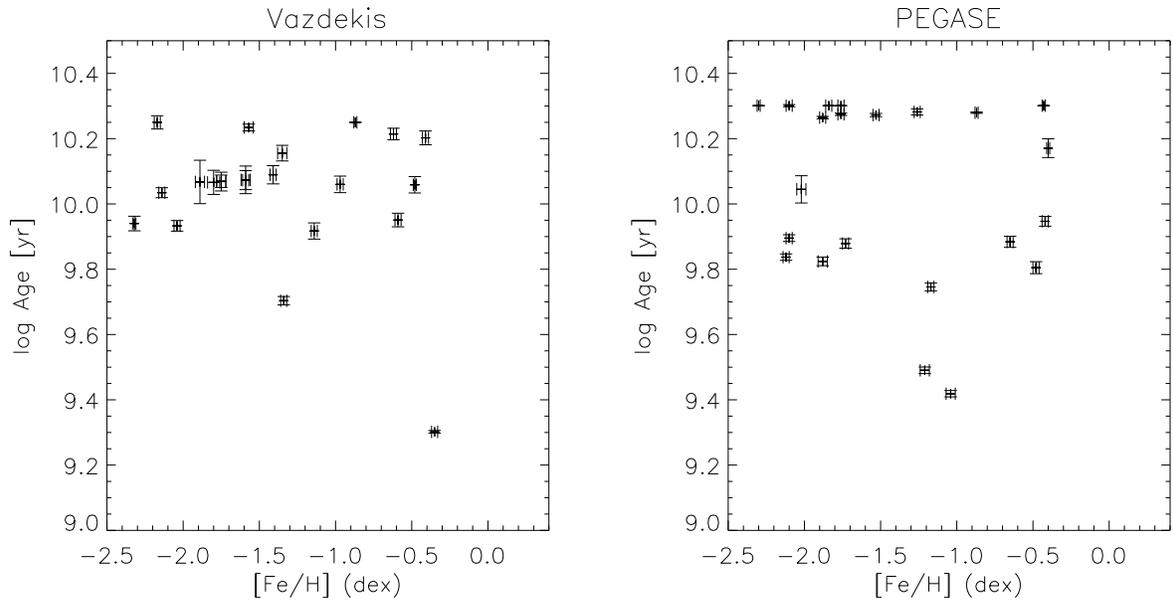}}}
  \caption{{\sc ULySS} fit results: Metallicities versus ages of our
    sample GCs derived with the spectroscopic fitting. The error bars
    for both the ages and metallicities are shown in the plots, but
    they are very small. Left: \citet{vaz} SSP models; Right: {\sc
    pegase-hr} SSP models.}
  \label{fig3}
\end{figure}

\begin{figure}
  \centering
  \includegraphics[scale=0.7]{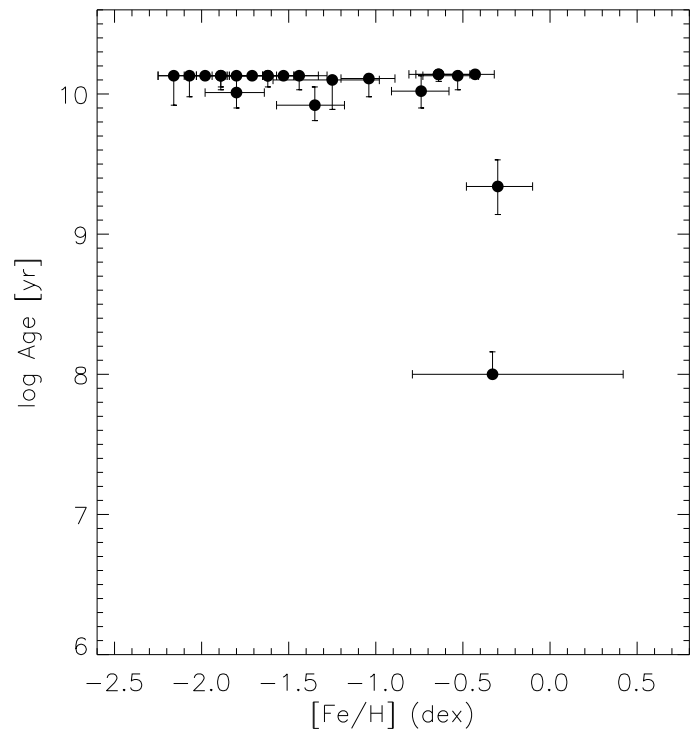}
  \caption{Same as Fig.~\ref{fig3}, but for the Thomas et al. models.}
  \label{fig4}
\end{figure}

\begin{figure}
  \centering
  \includegraphics[scale=0.7]{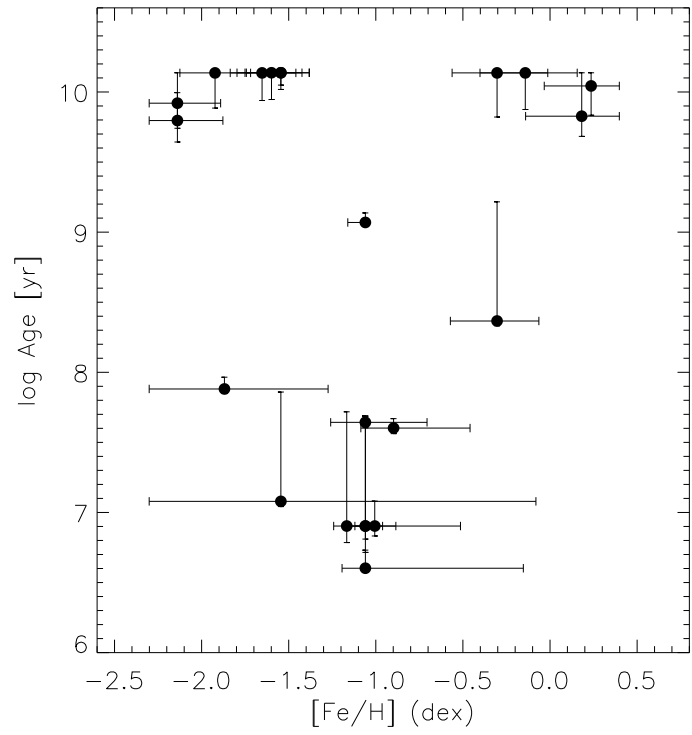}
  \caption{Same as Fig.~\ref{fig3}, but for the {\sc galev} models.}
  \label{fig5}
\end{figure}

\begin{figure}[!htbp]
  \centering
  \subfigure[Padova 1994 evolutionary tracks; Chabrier IMF]{
    \label{fig:first_sub}
    \includegraphics[scale=0.53]{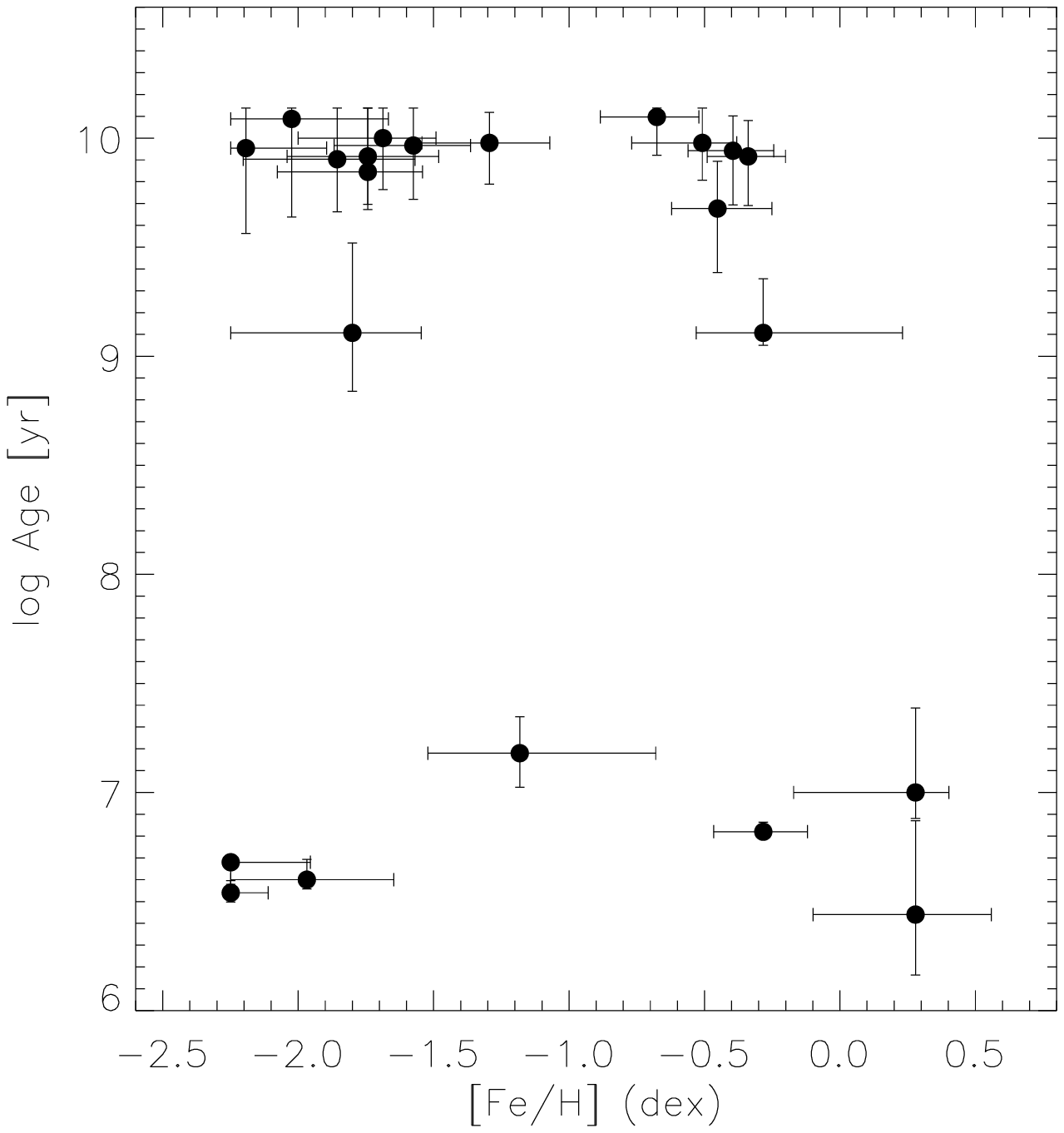}}
  %  \hspace{0.5in}
  \subfigure[Padova 1994 evolutionary tracks; Salpeter IMF]{
    \label{fig:second_sub}
    \includegraphics[scale=0.53]{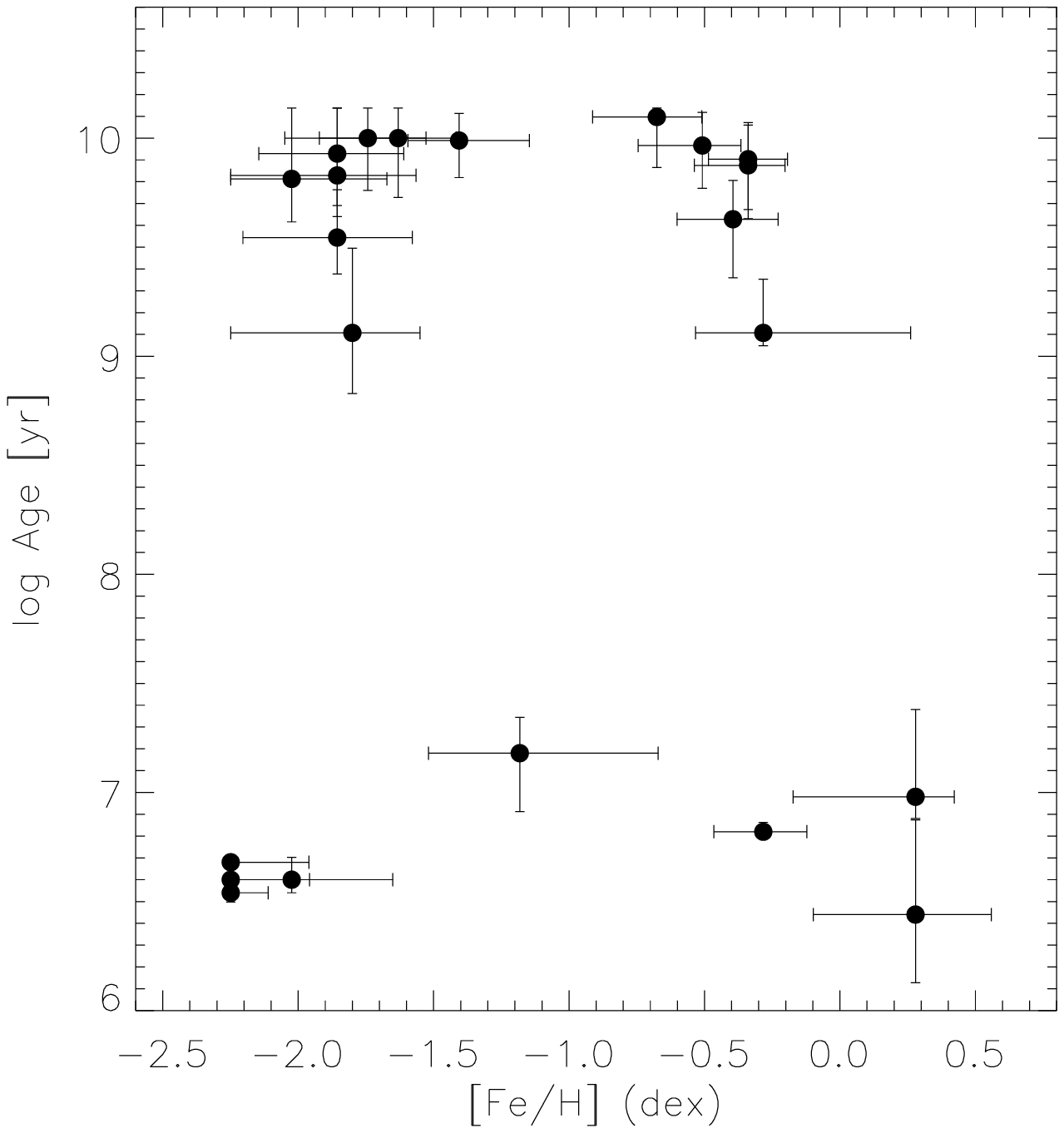}}
  %  \hspace{0.5in}
  \subfigure[Padova 2000 evolutionary tracks; Chabrier IMF]{
    \label{fig:third_sub}
    \includegraphics[scale=0.53]{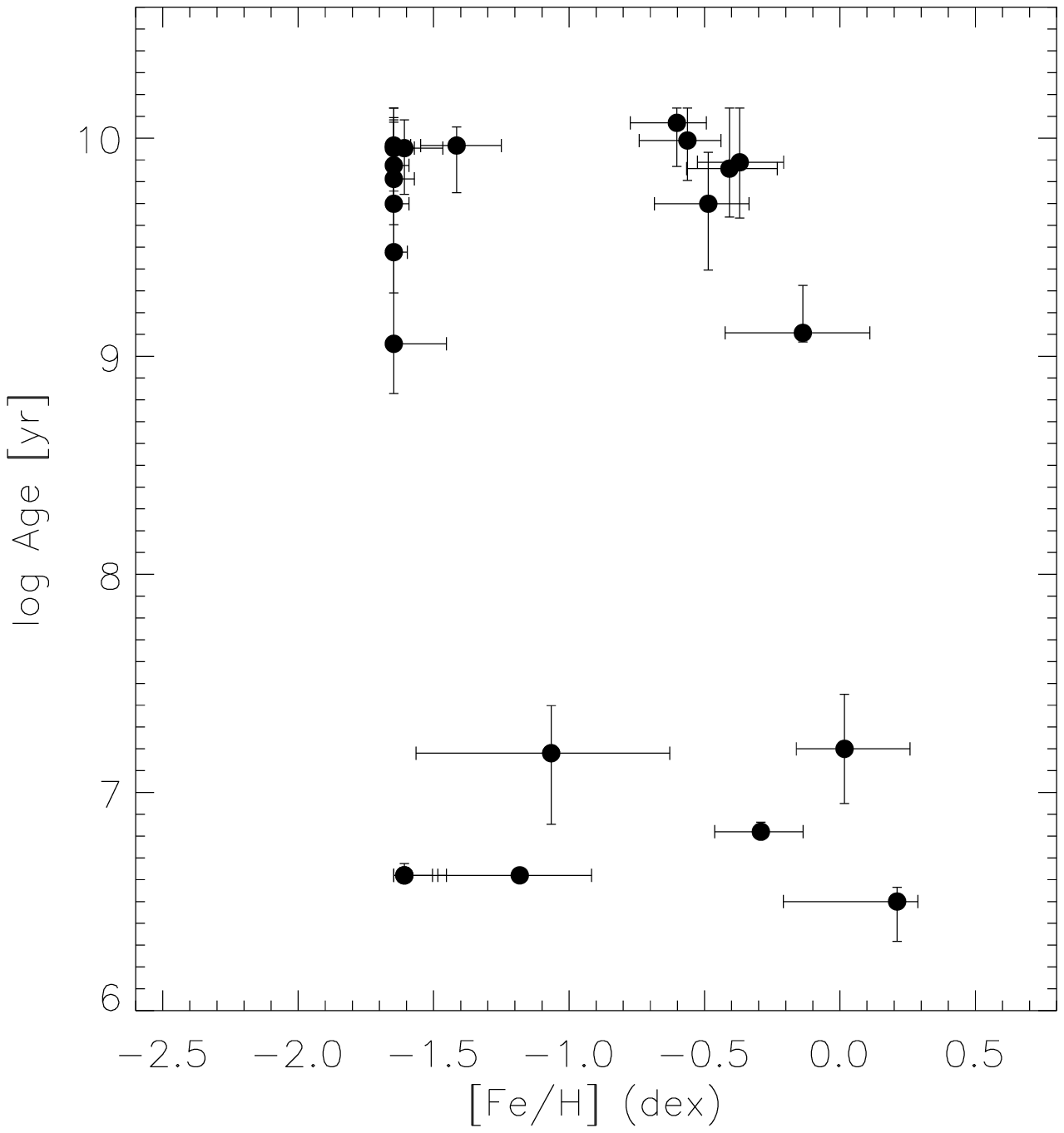}}
  %  \hspace{0.5in}
  \subfigure[Padova 2000 evolutionary tracks; Salpeter IMF]{
    \label{fig:fourth_sub}
    \includegraphics[scale=0.53]{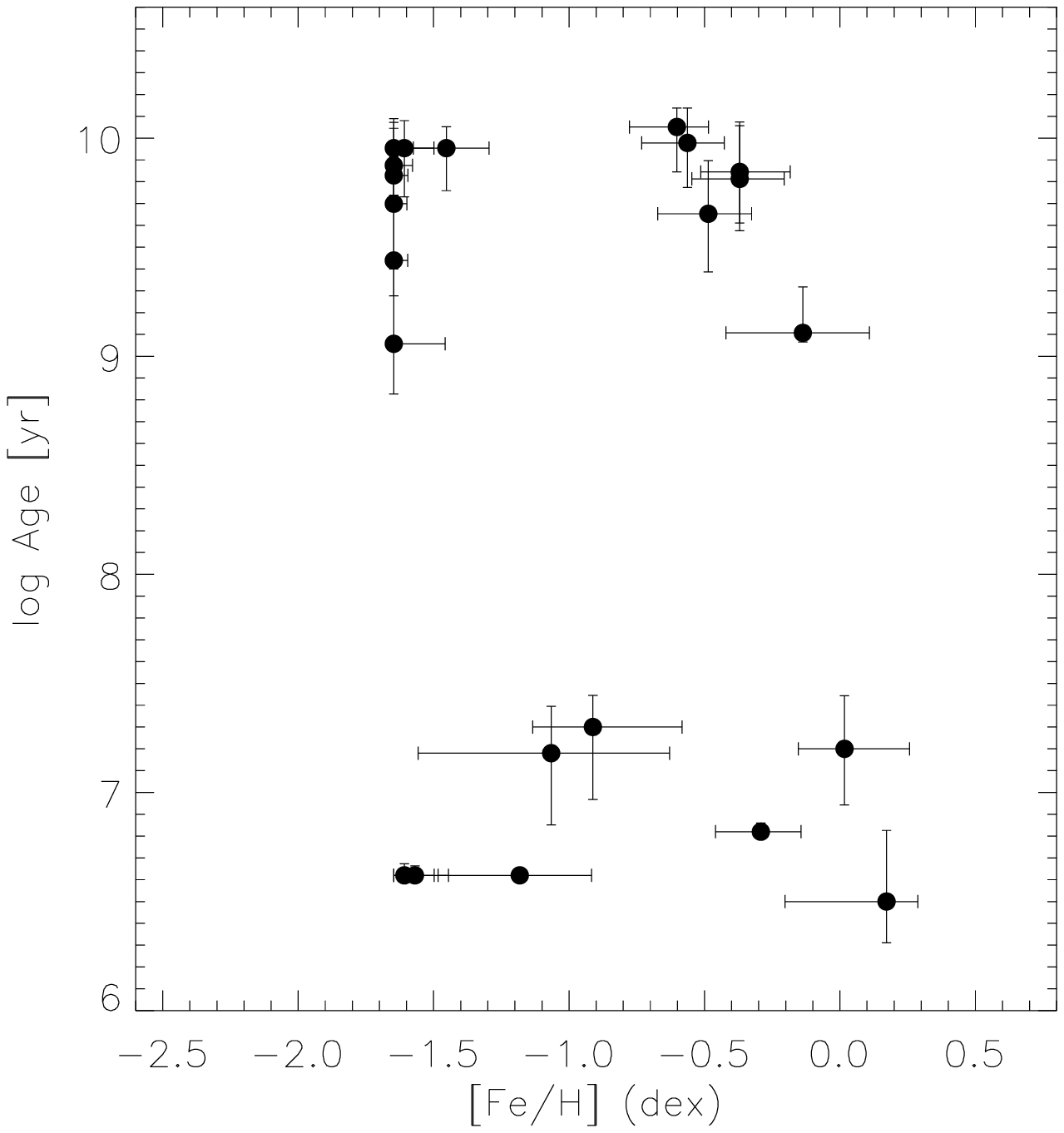}}
  \caption{Metallicities versus ages of our sample GCs derived only
    from Lick/IDS absorption-line index fits with the \citet{bc03}
    models, Padova 1994/2000 evolutionary tracks, and
    Chabrier/Salpeter IMFs.}
  \label{fig6}
\end{figure}

\begin{figure}[!htbp]
  \centering
  \subfigure[Padova 1994 evolutionary tracks; Chabrier IMF]{
    \label{fig:first_suba}
    \includegraphics[scale=0.53]{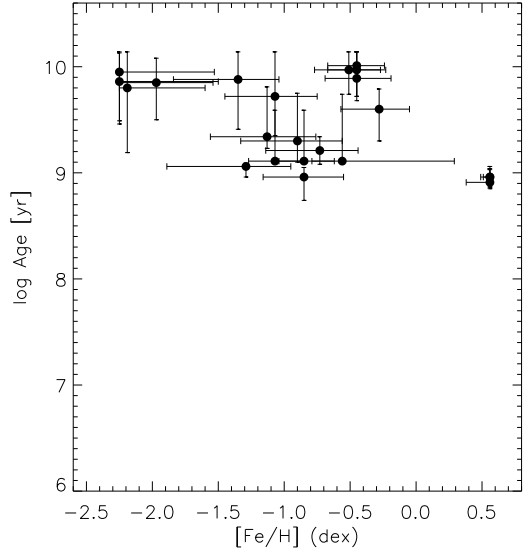}}
  %  \hspace{0.5in}
  \subfigure[Padova 1994 evolutionary tracks; Salpeter IMF]{
    \label{fig:second_suba}
    \includegraphics[scale=0.53]{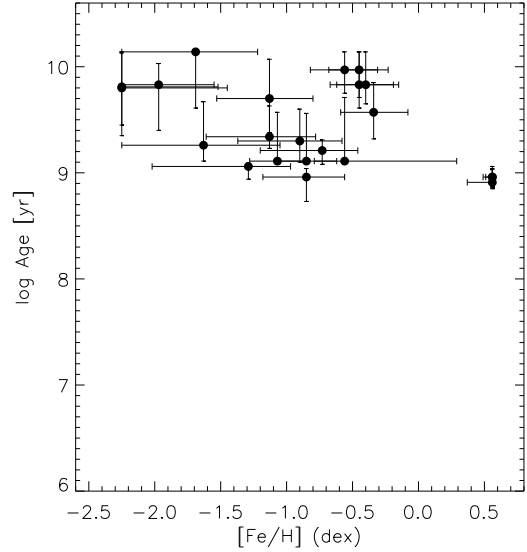}}
  %  \hspace{0.5in}
  \subfigure[Padova 2000 evolutionary tracks; Chabrier IMF]{
    \label{fig:third_suba}
    \includegraphics[scale=0.53]{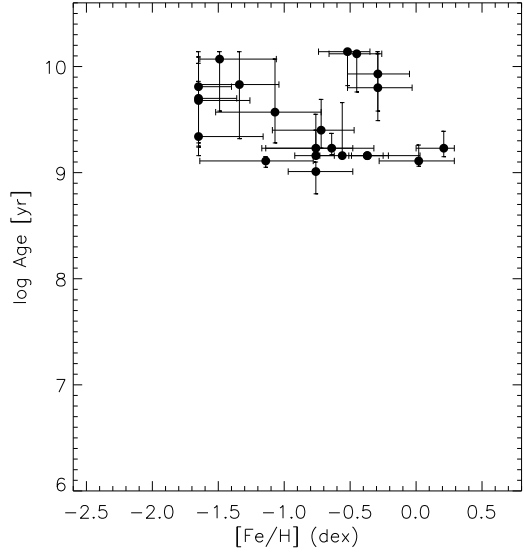}}
  %  \hspace{0.5in}
  \subfigure[Padova 2000 evolutionary tracks; Salpeter IMF]{
    \label{fig:fourth_suba}
    \includegraphics[scale=0.53]{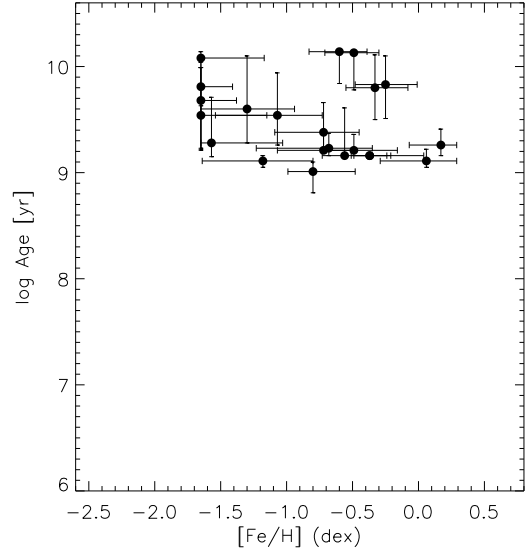}}
  \caption{Metallicities versus ages of our sample GCs derived only
    based on SED fitting with the \citet{bc03} models, Padova
    1994/2000 evolutionary tracks, and Chabrier/Salpeter IMFs.}
  \label{fig7}
\end{figure}

\begin{figure}[!htbp]
  \centering
  \subfigure[Padova 1994 evolutionary tracks; Chabrier IMF]{
    \label{fig:first_subb}
    \includegraphics[scale=0.53]{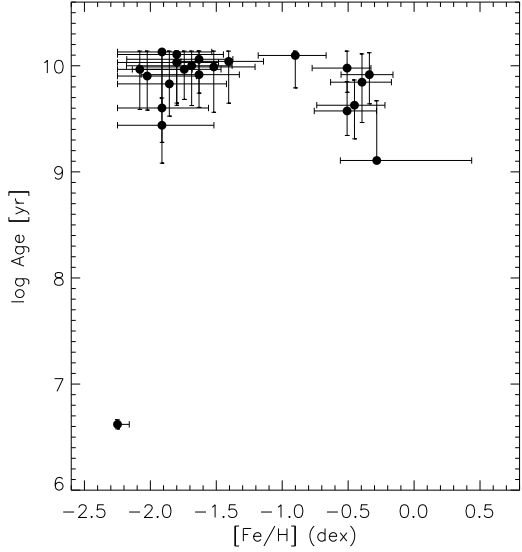}}
  %  \hspace{0.5in}
  \subfigure[Padova 1994 evolutionary tracks; Salpeter IMF]{
    \label{fig:second_subb}
    \includegraphics[scale=0.53]{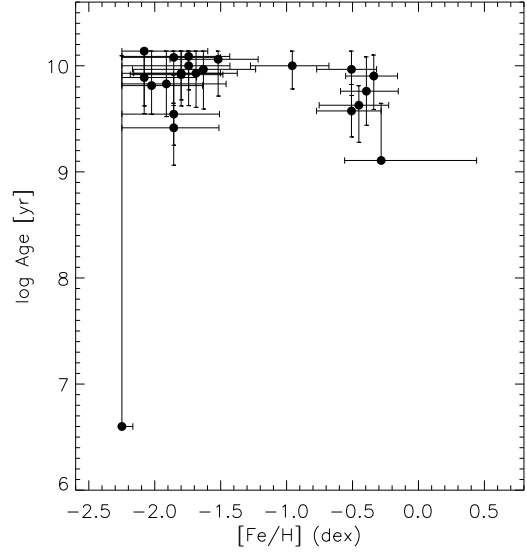}}
  %  \hspace{0.5in}
  \subfigure[Padova 2000 evolutionary tracks; Chabrier IMF]{
    \label{fig:third_subb}
    \includegraphics[scale=0.53]{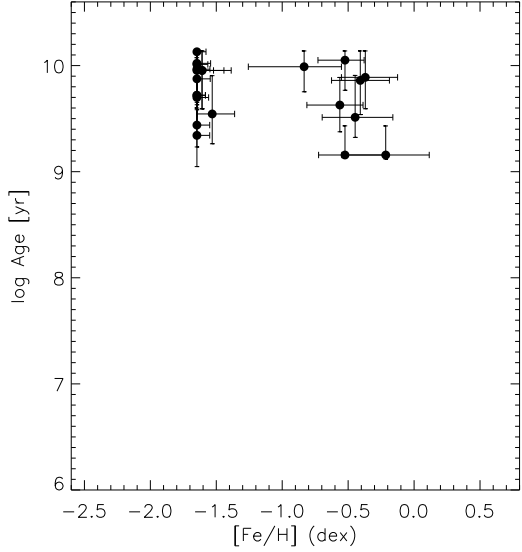}}
  %  \hspace{0.5in}
  \subfigure[Padova 2000 evolutionary tracks; Salpeter IMF]{
    \label{fig:fourth_subb}
    \includegraphics[scale=0.53]{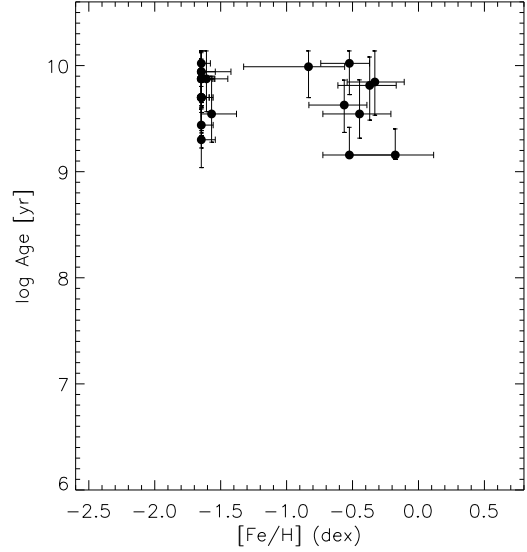}}
  \caption{Metallicities versus ages of our sample GCs derived from
    both the Lick/IDS absorption indices and the SED fits with the
    \citet{bc03} models, Padova 1994/2000 stellar evolutionary tracks,
    and Chabrier/Salpeter IMFs.}
  \label{fig8}
\end{figure}

\begin{figure}
  \centering
  \includegraphics[scale=0.67]{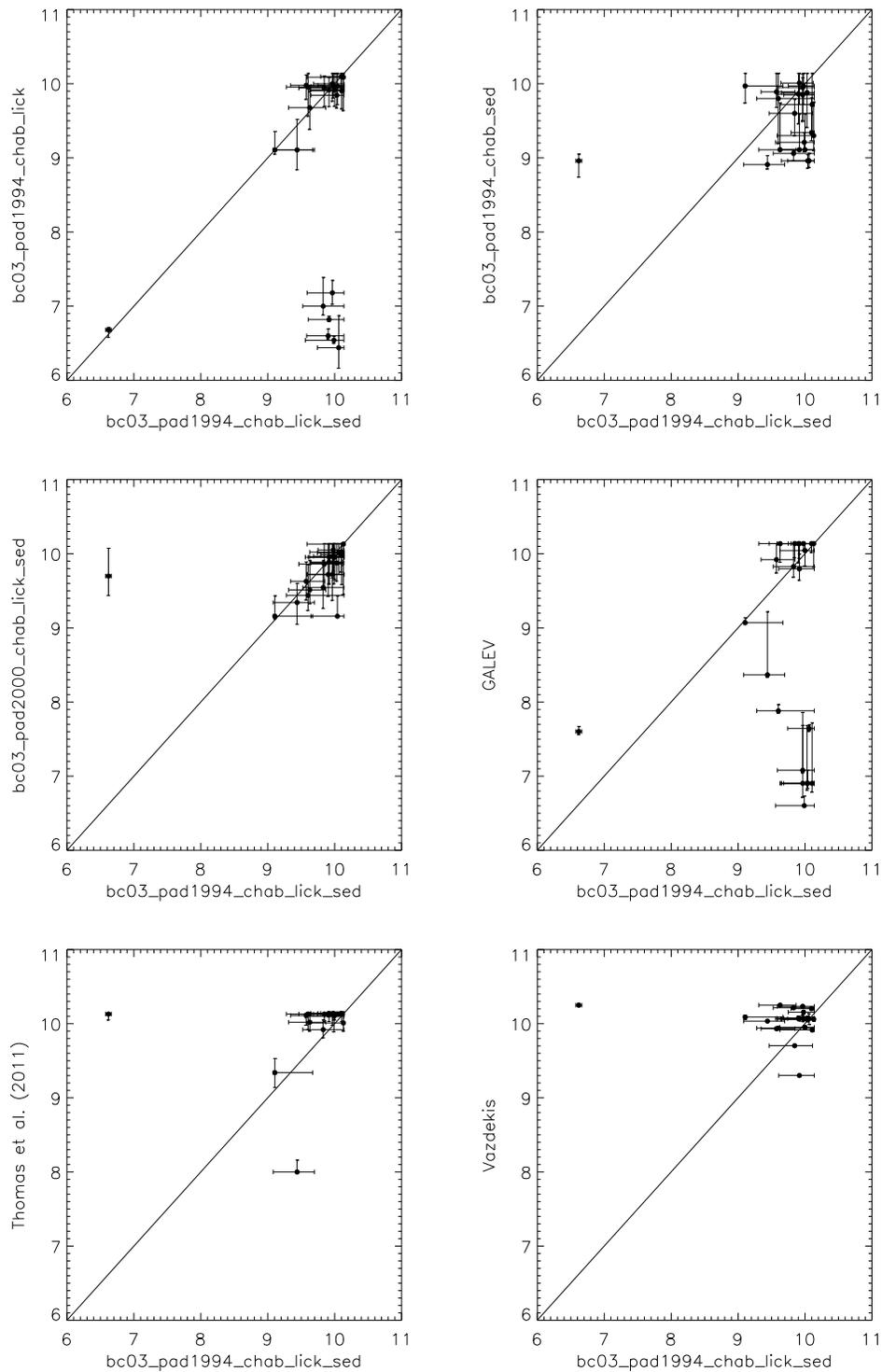}
  \caption{Comparisons of log( age ) [yr] derived from different
    models. All $x$ axes are the BC03 (Padova 1994 + \citet{chab03})
    combined fit results to the Lick indices and SEDs. Regarding the
    $y$ axes, top left: BC03 (Padova 1994 + \citet{chab03}) fits to
    only the Lick indices; top right: BC03 (Padova 1994 +
    \citet{chab03}) fits to only the SEDs; middle left: BC03 (Padova
    2000 + \citet{chab03}) combined fitting of Lick indices and SEDs;
    middle right: {\sc galev} model fitting to only the Lick indices;
    bottom left: \citet{tmj} fitting of only the Lick indices; bottom
    right: {\sc ULySS} models and Vazdekis full-spectrum fitting.}
  \label{fig9}
\end{figure}

\begin{figure}
  \centering
  \includegraphics[scale=0.67]{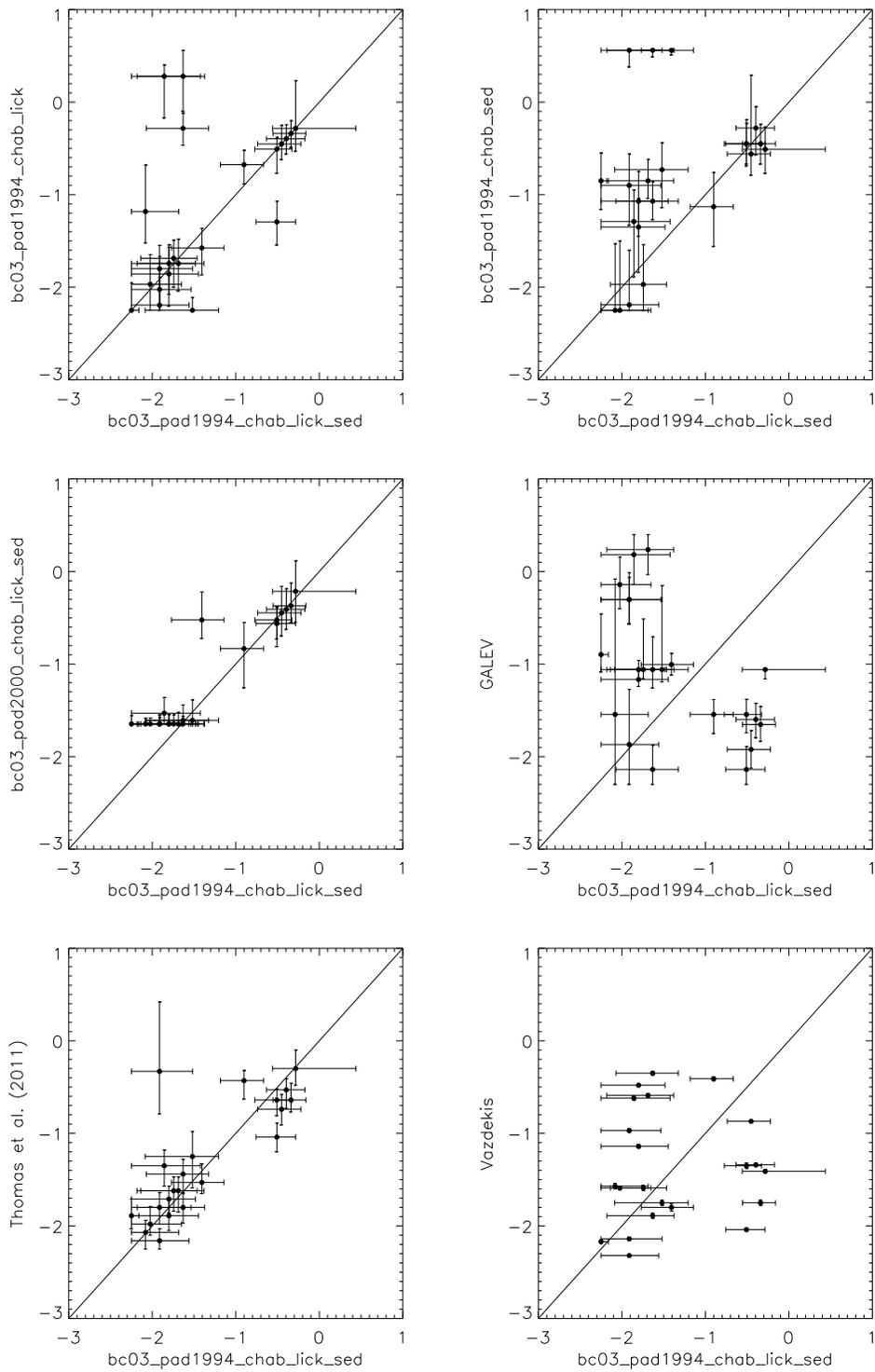}
  \caption{Comparisons of metallicities $\rm [Fe/H]$ derived from
    different models. Layout and panel coding: same as
    Fig.~\ref{fig9}.}
  \label{fig10}
\end{figure}

For all sample clusters, we measured all 20 Lick absorption-line
indices \citep{w94a,wo97} and fitted the ages, metallicities, and
radial velocities with $\rm EZ\_Ages$. We also performed full-spectrum
fitting with the {\sc ULySS} code \citep{kol09}, adopting the
\citet{vaz} and {\sc pegase-hr} SSP models, to obtain the ages,
metallicities, and radial velocities. Similarly to
\citet{sha,fan11,fan12}, we applied $\chi^2_{\rm min}$ fitting to the
Lick absorption-line indices with the updated \citet{tmj}, {\sc
  galev}, and BC03 \citep{bc03} SSP models for Padova 1994/2000
stellar evolutionary tracks and \citet{chab03} or \citet{sal}
IMFs. The results show that most HGCs are old ($>10$ Gyr) and
metal-poor. For {\sc ULySS} and the \citet{vaz} SSP models, $\rm
age_{mean}=11.7$ Gyr (r.m.s. = 3.9 Gyr) and $\rm [Fe/H]_{mean}=-1.37$
dex (r.m.s. = 0.60 dex); for the {\sc pegase-hr} models, $\rm
age_{mean}=12.9$ Gyr (r.m.s. = 6.4 Gyr) and $\rm [Fe/H]_{mean}=-1.41$
dex (r.m.s. = 0.62 dex), which indicates that these halo star clusters
were born during the earliest stages of the galaxy's formation.
However, also note that there are several clusters with relatively
high metallicities and younger ages (see Figs
\ref{fig3}--\ref{fig6}). These clusters cloud have had their origins
in disrupted dwarf galaxies accreted by M31 \citep{cbq}.

For comparison, we collected cluster photometry in the $UBVRIHJK$
bands from the RBC v.5 and fitted the SEDs of our sample GCs. In
addition, we fitted a combination of SEDs and Lick absorption-line
indices, and we found that the results improved significantly. The
fits' uncertainties became smaller and the results were more
reliable. Therefore, we conclude that fits to a combination of SEDs
and Lick indices are significantly better than SED or Lick-index fits
alone.

We compared the ages derived from fitting with different models and
methods. The fit results from the Lick indices only agree well with
the combined fit results if we adopt the BC03 models, the Padova 1994
tracks, and a \citet{chab03} IMF, except for a few outliers. The fits
to only the SEDs agree well with the combined fits. The agreement is
even better than for the Lick indices only. This indicates that any
differences between the Padova 1994 and 2000 evolutionary tracks are
insignificant as regards the resulting ages. A few of the `old' star
clusters returned by the BC03 models are considered `young' based on
the {\sc galev} models, the reasons for which were explained by
\citet{fd12}. Both the results from the \citet{tmj} models and the
{\sc ULySS} fits with the \citet{vaz} SSP models agree well with the 
results from the BC03 combined fits to the Lick indices and SEDs.

The metallicities resulting from the different models and fitting
methods were also compared. The BC03 fit results to only the Lick
indices agree well with the combined fit results, expect for a few
outliers. However, the BC03 fit results to only the SEDs do not agree
as well with the combined fits owing to the large scatter, compared
with fits to only the Lick indices. The fit results using the BC03
models, Padova 1994/2000 tracks, and a \citet{chab03} IMF agree well
with each other, except for the low-metallicity range because of
differences between the models. The metallicities delivered by the
{\sc galev} and {\sc ULySS} models do not agree as well with the
combined fit results based on the BC03 models as those for other
models. The results from the \citet{tmj} models agree much better with
the combined fits from the BC03 models than with those from other
models. In future research, we may consider more models for the
comparison, e.g., binary-star SSP models \citep{li08,li13}. Binary
stars effect the color and Lick indices by a few percent. Although
this is smaller than the systematic errors associated with the models,
it makes the results fainter and bluer.

Although a large number of studies have already explored the
properties of the M31 halo star clusters, our understanding of the
interaction of M31 and M33 with our Milky Way is still far from clear,
because the cluster sample is incomplete, especially for fainter
luminosities. We aim at obtaining additional spectroscopic
observations to enlarge the halo star cluster sample of M31 and M33 to
remedy this situation.

\acknowledgements

This research was supported by the National Natural Science Foundation
of China (NFSC) through grants 11003021, 11373003, 11373010, 11563002,
and U1631102, and the National Key Basic Research Program of China
(973 Program), grant 2015CB857002. Z.F. acknowledges a Young
Researcher Grant of the National Astronomical Observatories, Chinese
Academy of Sciences.

\appendix		   %%appendicial material is supported

\clearpage
\pagestyle{empty}
\begin{deluxetable}{lccrrrr}
  \tablecolumns{6} \tablewidth{0pc} \tablecaption{Observations of our Sample GCs.\label{t1.tab}}
%%Please Capitalize the First Letter of Each Notional Word in table's
%%caption
  \tablehead{\colhead{ID} & \colhead{R.A.} & \colhead{Dec.} &
    \colhead{$r_{\rm p}$} & \colhead{$V$} &\colhead{Exp} & \colhead{Date}\\
    & (J2000) & (J2000) & (kpc) & (mag) & (s)& (mm/dd/year) \\}
    \startdata
 H9	   & 00:34:17.223 & +37:30:43.42 & 56.08 & 17.72 & 600$\times$2 &  10/31/2010 \\
 MCGC5-H10 & 00:35:59.700 & +35:41:03.60 & 78.68 & 16.13 & 480		&  10/31/2010 \\
 SK001A    & 00:36:33.523 & +40:39:45.04 & 17.97 & 17.66 & 900		&  11/04/2011 \\
 B423	   & 00:37:56.662 & +40:57:35.96 & 13.04 & 17.83 & 900		&  11/04/2011 \\
 B298-MCGC6& 00:38:00.300 & +40:43:56.10 & 14.25 & 16.48 & 480		&  11/01/2010 \\
 H12	   & 00:38:03.870 & +37:44:00.70 & 49.93 & 16.54 & 480		&  10/31/2010 \\
 B167D	   & 00:38:22.480 & +41:54:35.06 & 14.18 & 17.90 & 720$\times$2 &  10/31/2010 \\
 B309-G031 & 00:39:24.606 & +40:14:29.12 & 16.48 & 17.44 & 600$\times$2 &  10/31/2010 \\
 B436	   & 00:39:30.668 & +40:18:20.50 & 15.59 & 18.22 & 600$\times$2 &  10/31/2010 \\
 H15	   & 00:40:13.214 & +35:52:37.02 & 74.22 & 17.99 & 900		&  10/31/2010 \\
 B006-G058 & 00:40:26.488 & +41:27:26.71 &  6.43 & 15.50 & 300		&  10/31/2010 \\
 H16	   & 00:40:37.800 & +39:45:29.90 & 21.37 & 17.54 & 900$\times$2 &  10/31/2010 \\
 B017-G070 & 00:40:48.716 & +41:12:07.21 &  5.03 & 15.95 & 360		&  11/01/2010 \\
 B339-G077 & 00:41:00.709 & +39:55:54.21 & 18.82 & 16.88 & 600		&  11/01/2010 \\
 B361-G255 & 00:43:57.100 & +40:14:01.25 & 14.50 & 16.95 & 480		&  11/01/2010 \\
 G260	   & 00:44:00.857 & +42:34:48.48 & 18.21 & 16.92 & 900$\times$2 &  11/02/2010 \\
 SK104A    & 00:45:44.280 & +41:57:27.40 & 12.13 & 17.98 & 600$\times$2 &  10/31/2010 \\
 B396-G335 & 00:47:25.158 & +40:21:42.14 & 17.33 & 17.32 & 600$\times$2 &  10/31/2010 \\
 G339-BA30 & 00:47:50.215 & +43:09:16.43 & 28.83 & 17.20 & 480$\times$2 &  10/31/2010 \\
 B402-G346 & 00:48:36.046 & +42:01:34.54 & 18.20 & 17.27 & 900$\times$2 &  11/02/2010 \\
 B337D	   & 00:49:11.209 & +41:07:21.06 & 16.70 & 18.13 & 720$\times$2 &  10/31/2010 \\
 H22	   & 00:49:44.690 & +38:18:37.40 & 44.47 & 17.03 & 900$\times$2 &  11/02/2010 \\
\enddata
\end{deluxetable}

\clearpage
\pagestyle{empty}
\begin{deluxetable}{lrrr}
  \tablecolumns{3} \tablewidth{0pc} \tablecaption{Radial Velocities,
    $V_r$, of our Sample GCs as well as Previously Published
    Measurements.
    \label{t2.tab}}
%%Please Capitalize the First Letter of Each Notional Word in table's
  %%caption
  \tablehead{\colhead{ID} & \colhead{\citet{vaz}} & \colhead{PEGASE-HR} & \colhead{RBC v.5}\\}
  \startdata
  H9	    & $-603\pm 2$ & $-594\pm 2$ & ... ...  \\
  MCGC5-H10 & $-533\pm 3$ & $-537\pm 2$ & $-358\pm 2$  \\
  SK001A    &	$45\pm 2$ & $54\pm 2$ & $-240\pm47$  \\
  B423	    &  $-296\pm5$ & $-349\pm5$ & $-215\pm46$  \\
  B298-MCGC6$^1$& $-551\pm3$ & $-540\pm2$ &$-539\pm12$	\\
  H12	    & $-651\pm9$ & $-655\pm13$ & ... ...  \\
  B167D     & $-379\pm 3$ & $-399\pm 4$ & $-196\pm15$	  \\
  B309-G031 & $-555\pm 3$ & $-527\pm 2$ &$-480\pm26$  \\
  B436	    & $-654\pm1$ & $-681\pm 2$ & $-488\pm18$  \\
  H15	    & $-138\pm2$ & $-36\pm1$  & ... ... \\
  B006-G058 & $-319\pm 1$ & $-321\pm 1$ & $-237\pm 1$ \\
  H16	    & $-673\pm2$ & $-643\pm2$ & ... ...  \\
  B017-G070 &  $-548\pm 1$ &  $-543\pm 1$ & $-514\pm 8$ \\
  G339-BA30 & $-244\pm 1$ & $-234\pm 1$ &  $33\pm30$  \\
  B361-G255 & $-353\pm 1$ & $-351\pm 2$ & $-330\pm26$  \\
  G260	    & $-218\pm1$ & $-219\pm 1$ &  $16\pm26$  \\
  SK104A    &  $-195\pm 1$ & $-194\pm1$ & $-301\pm17$ \\
  B396-G335 & $-581\pm16$ &$-528\pm 16$ &  $-561\pm30$	\\
  G339-BA30 & $-225\pm6$ & $-247\pm60$ &  $33\pm30$  \\
  B402-G346 & $-351\pm120$ & $-332\pm1$  & $-488\pm26$	\\
  B337D     & $-422\pm 3$ & $-397\pm 3$ & $-222\pm23$  \\
  H22	    & $-333\pm 2$ & $-329\pm 2$ & ... ...  \\
  \enddata
\end{deluxetable}

\clearpage

\pagestyle{empty}
\begin{deluxetable}{lrrrr}
  \tablecolumns{3} \tablewidth{0pc} \tablecaption{Ages and
    Metallicities Derived from the Lick Absorption-Line Indices Fitted
    with {\sc ULySS}.\label{t3.tab}} \tablehead{\colhead{Name} &
    \colhead{log $\rm Age_{Vazdekis}$} & \colhead{log $\rm
      Age_{PEGASE-HR}$} & \colhead{$\rm [Fe/H]_{Vazdekis}$} &
    \colhead{$\rm [Fe/H]_{PEGASE-HR}$}\\ \colhead{} & \colhead{(yr)} &
    \colhead{(yr)} & \colhead{(dex)} & \colhead{(dex)}} \startdata
           H9 & $ 10.069\pm 0.029$ & $  9.879\pm 0.014$ & $ -1.750\pm 0.030$ & $ -1.730\pm 0.020$ \\
  MCGC5-H10 & $ 10.156\pm 0.024$ & $ 10.301\pm 0.000$ & $ -1.350\pm 0.030$ & $ -1.840\pm 0.020$ \\
     SK001A & $ 10.202\pm 0.021$ & $ 10.171\pm 0.029$ & $ -0.410\pm 0.020$ & $ -0.400\pm 0.010$ \\
       B423 & $ 10.234\pm 0.010$ & $ 10.272\pm 0.005$ & $ -1.570\pm 0.030$ & $ -1.530\pm 0.020$ \\
 B298-MCGC6 & $ 10.034\pm 0.015$ & $  9.837\pm 0.010$ & $ -2.140\pm 0.020$ & $ -2.120\pm 0.020$ \\
        H12 & $ 10.074\pm 0.042$ & $ 10.282\pm 0.009$ & $ -1.590\pm 0.030$ & $ -1.260\pm 0.020$ \\
      B167D & $ 10.069\pm 0.019$ & $ 10.275\pm 0.004$ & $ -1.750\pm 0.030$ & $ -1.760\pm 0.020$ \\
  B309-G031 & $ 10.073\pm 0.029$ & $ 10.265\pm 0.004$ & $ -1.590\pm 0.020$ & $ -1.880\pm 0.020$ \\
       B436 & $  9.302\pm 0.004$ & $ 10.301\pm 0.003$ & $ -0.350\pm 0.020$ & $ -2.100\pm 0.020$ \\
        H15 & $  9.940\pm 0.022$ & $  9.490\pm 0.010$ & $ -2.320\pm 0.010$ & $ -1.210\pm 0.030$ \\
  B006-G058 & $  9.951\pm 0.021$ & $  9.805\pm 0.018$ & $ -0.590\pm 0.020$ & $ -0.480\pm 0.020$ \\
        H16 & $ 10.250\pm 0.020$ & $ 10.301\pm 0.000$ & $ -2.170\pm 0.020$ & $ -2.300\pm 0.010$ \\
  B017-G070 & $ 10.060\pm 0.025$ & $ 10.280\pm 0.002$ & $ -0.970\pm 0.020$ & $ -0.870\pm 0.010$ \\
  B339-G077 & $ 10.214\pm 0.018$ & $  9.947\pm 0.015$ & $ -0.620\pm 0.020$ & $ -0.420\pm 0.020$ \\
  B361-G255 & $ 10.089\pm 0.028$ & $  9.746\pm 0.012$ & $ -1.410\pm 0.020$ & $ -1.170\pm 0.020$ \\
       G260 & $  9.704\pm 0.013$ & $  9.419\pm 0.010$ & $ -1.340\pm 0.020$ & $ -1.040\pm 0.030$ \\
     SK104A & $ 10.059\pm 0.025$ & $ 10.301\pm 0.000$ & $ -0.480\pm 0.010$ & $ -0.430\pm 0.010$ \\
  B396-G335 & $ 10.067\pm 0.066$ & $ 10.045\pm 0.042$ & $ -1.890\pm 0.030$ & $ -2.020\pm 0.030$ \\
  G339-BA30 & $ 10.066\pm 0.037$ & $  9.824\pm 0.013$ & $ -1.800\pm 0.040$ & $ -1.880\pm 0.030$ \\
  B402-G346 & $ 10.250\pm 0.000$ & $  9.884\pm 0.017$ & $ -0.870\pm 0.010$ & $ -0.650\pm 0.020$ \\
      B337D & $  9.917\pm 0.025$ & $ 10.301\pm 0.000$ & $ -1.140\pm 0.020$ & $ -1.760\pm 0.020$ \\
        H22 & $  9.933\pm 0.016$ & $  9.895\pm 0.010$ & $ -2.040\pm 0.020$ & $ -2.100\pm 0.020$ \\
 \enddata
\end{deluxetable}

\pagestyle{empty}
\begin{deluxetable}{lrrrr}
  \tablecolumns{3} \tablewidth{0pc} \tablecaption{Ages and
    Metallicities Derived from the Lick Absorption-Line Indices Fitted
    with $\rm EZ\_Ages$.\label{t4.tab}} \tablehead{\colhead{Name} &
    \colhead{log $\rm Age_{EZ\_Ages}$} & \colhead{$\rm
      [Fe/H]_{EZ\_Ages}$} & \colhead{$\rm [Fe/H]_{[MgFe]}$} &
    \colhead{$\rm [Fe/H]_{\langle Fe \rangle}$} \\ \colhead{} &
    \colhead{(Gyr)} & \colhead{(dex)} & \colhead{(dex)} &
    \colhead{(dex)}} \startdata          H9 &  99.99 & $ 99.99$ & $ -1.93$ & $ -2.50$ \\
  MCGC5-H10 &  99.99 & $ 99.99$ & $ -1.49$ & $ -1.50$ \\
     SK001A &  12.21 & $ -0.90$ & $ -0.35$ & $ -0.90$ \\
       B423 &  99.99 & $ 99.99$ & $ -2.23$ & $ -2.95$ \\
 B298-MCGC6 &  99.99 & $ 99.99$ & $ -1.74$ & $ -1.99$ \\
        H12 &  99.99 & $ 99.99$ & $ -1.86$ & $ -2.03$ \\
      B167D &   4.36 & $ -1.12$ & $ -1.56$ & $ -1.28$ \\
  B309-G031 &   3.99 & $ -0.84$ & $ -0.97$ & $ -1.06$ \\
       B436 &   2.33 & $ -0.76$ & $ -1.57$ & $ -1.08$ \\
        H15 &   8.45 & $ -1.00$ & $ -1.24$ & $ -1.08$ \\
  B006-G058 &  99.99 & $ 99.99$ & $ -0.42$ & $ -0.62$ \\
        H16 &  99.99 & $ 99.99$ & $ -0.80$ & $ -0.81$ \\
  B017-G070 &   7.03 & $ -0.77$ & $ -0.85$ & $ -0.89$ \\
  B339-G077 &  99.99 & $ 99.99$ & $ -0.57$ & $ -0.78$ \\
  B361-G255 &   2.09 & $ -0.57$ & $ -0.91$ & $ -0.95$ \\
       G260 &  99.99 & $ 99.99$ & $ -1.50$ & $ -1.59$ \\
     SK104A &  99.99 & $ 99.99$ & $ -0.44$ & $ -0.60$ \\
  B396-G335 &  99.99 & $ 99.99$ & $ -2.25$ & $ -2.73$ \\
  G339-BA30 &  99.99 & $ 99.99$ & $ -1.54$ & $ -1.48$ \\
  B402-G346 &   3.18 & $ -0.18$ & $ -0.43$ & $ -0.39$ \\
      B337D &   3.90 & $ -1.09$ & $ -1.18$ & $ -1.28$ \\
        H22 &  99.99 & $ 99.99$ & $ -1.89$ & $ -2.08$ \\
 \enddata
\end{deluxetable}

\clearpage

\pagestyle{empty}
\begin{deluxetable}{lrrr}
  \tablecolumns{4} \tablewidth{0pc} \tablecaption{Ages and
    Metallicities Derived from the Lick Absorption-Line Indices Fitted
    with the \citet{tmj} models.\label{t5.tab}}
  \tablehead{\colhead{ID} & \colhead{log Age} & \colhead{$\rm [Z/H]$}
    & \colhead{$\rm [\alpha/Fe]$}\\ & \colhead{[yr]} & \colhead{(dex)}
    & \colhead{(dex)}} \startdata
           H9 & $10.134^{+ 0.003}_{-0.082}$ & $-1.620^{+ 0.153}_{-0.227}$ & $ 0.300^{+ 0.200}_{-0.600}$ \\
  MCGC5-H10 & $10.134^{+ 0.003}_{-0.017}$ & $-1.710^{+ 0.139}_{-0.146}$ & $ 0.500^{+ 0.000}_{-0.600}$ \\
     SK001A & $10.137^{+ 0.000}_{-0.029}$ & $-0.432^{+ 0.106}_{-0.195}$ & $ 0.500^{+ 0.000}_{-0.227}$ \\
       B423 & $10.134^{+ 0.003}_{-0.020}$ & $-1.800^{+ 0.155}_{-0.172}$ & $ 0.500^{+ 0.000}_{-0.367}$ \\
 B298-MCGC6 & $10.134^{+ 0.003}_{-0.212}$ & $-2.160^{+ 0.134}_{-0.090}$ & $ 0.480^{+ 0.020}_{-0.780}$ \\
        H12 & $10.134^{+ 0.003}_{-0.027}$ & $-1.980^{+ 0.191}_{-0.122}$ & $ 0.500^{+ 0.000}_{-0.800}$ \\
      B167D & $10.134^{+ 0.003}_{-0.079}$ & $-1.890^{+ 0.178}_{-0.141}$ & $-0.300^{+ 0.800}_{ 0.000}$ \\
  B309-G031 & $10.134^{+ 0.003}_{-0.019}$ & $-1.620^{+ 0.151}_{-0.215}$ & $ 0.500^{+ 0.000}_{-0.800}$ \\
       B436 & $ 9.342^{+ 0.189}_{-0.198}$ & $-0.297^{+ 0.203}_{-0.181}$ & $-0.150^{+ 0.367}_{-0.150}$ \\
        H15 & $ 8.000^{+ 0.160}_{ 0.000}$ & $-0.330^{+ 0.750}_{-0.462}$ & $-0.300^{+ 0.128}_{ 0.000}$ \\
  B006-G058 & $10.137^{+ 0.000}_{-0.052}$ & $-0.636^{+ 0.178}_{-0.126}$ & $-0.030^{+ 0.306}_{-0.270}$ \\
        H16 & $10.104^{+ 0.033}_{-0.214}$ & $-1.248^{+ 0.274}_{-0.343}$ & $-0.300^{+ 0.592}_{ 0.000}$ \\
  B017-G070 & $10.107^{+ 0.030}_{-0.130}$ & $-1.044^{+ 0.151}_{-0.158}$ & $ 0.500^{+ 0.000}_{-0.438}$ \\
  B339-G077 & $10.134^{+ 0.003}_{-0.098}$ & $-0.534^{+ 0.117}_{-0.196}$ & $ 0.180^{+ 0.320}_{-0.315}$ \\
  B361-G255 & $10.134^{+ 0.003}_{-0.097}$ & $-1.440^{+ 0.156}_{-0.203}$ & $ 0.500^{+ 0.000}_{-0.642}$ \\
       G260 & $ 9.924^{+ 0.129}_{-0.112}$ & $-1.350^{+ 0.174}_{-0.216}$ & $ 0.500^{+ 0.000}_{-0.577}$ \\
     SK104A & $10.137^{+ 0.000}_{-0.018}$ & $-0.636^{+ 0.125}_{-0.168}$ & $-0.300^{+ 0.322}_{ 0.000}$ \\
  B396-G335 & $10.130^{+ 0.006}_{-0.153}$ & $-2.070^{+ 0.130}_{-0.180}$ & $-0.300^{+ 0.800}_{ 0.000}$ \\
  G339-BA30 & $10.134^{+ 0.003}_{-0.096}$ & $-1.890^{+ 0.183}_{-0.162}$ & $ 0.000^{+ 0.500}_{-0.300}$ \\
  B402-G346 & $10.017^{+ 0.107}_{-0.120}$ & $-0.738^{+ 0.158}_{-0.169}$ & $-0.270^{+ 0.456}_{-0.030}$ \\
      B337D & $10.134^{+ 0.003}_{-0.020}$ & $-1.530^{+ 0.198}_{-0.125}$ & $-0.300^{+ 0.594}_{ 0.000}$ \\
        H22 & $10.009^{+ 0.122}_{-0.107}$ & $-1.800^{+ 0.161}_{-0.181}$ & $ 0.210^{+ 0.290}_{-0.510}$ \\
 \enddata
\end{deluxetable}

\pagestyle{empty}
\begin{deluxetable}{lrr}
  \tablecolumns{3} \tablewidth{0pc} \tablecaption{Ages and
    Metallicities Derived from the Lick Absorption-Line Indices Fitted
    with the {\sc galev} Models.\label{t6.tab}}
  \tablehead{\colhead{ID} & \colhead{log Age} & \colhead{$\rm [Fe/H]$}
    \\ & \colhead{[yr]} & \colhead{(dex)}} \startdata
           H9   & $10.043^{+ 0.094}_{-0.209}$ & $ 0.236^{+ 0.162}_{-0.269}$ \\
  MCGC5-H10   & $ 6.903^{+ 0.780}_{-0.093}$ & $-1.060^{+ 0.099}_{-0.101}$ \\
     SK001A   & $10.137^{+ 0.000}_{-0.119}$ & $-1.545^{+ 0.162}_{-0.206}$ \\
       B423   & $ 7.643^{+ 0.048}_{-0.039}$ & $-1.060^{+ 0.354}_{-0.199}$ \\
 B298-MCGC6   & $ 7.881^{+ 0.084}_{-0.033}$ & $-1.869^{+ 0.595}_{-0.432}$ \\
        H12   & $10.136^{+ 0.001}_{-0.261}$ & $-0.142^{+ 0.298}_{-0.260}$ \\
      B167D   & $ 7.602^{+ 0.067}_{-0.040}$ & $-0.898^{+ 0.439}_{-0.188}$ \\
  B309-G031   & $ 6.903^{+ 0.782}_{-0.188}$ & $-1.060^{+ 0.546}_{-0.099}$ \\
       B436   & $ 9.069^{+ 0.068}_{-0.003}$ & $-1.060^{+ 0.001}_{-0.101}$ \\
        H15   & $ 8.366^{+ 0.850}_{-0.034}$ & $-0.304^{+ 0.240}_{-0.268}$ \\
  B006-G058   & $10.136^{+ 0.001}_{-0.197}$ & $-1.653^{+ 0.194}_{-0.183}$ \\
        H16   & $ 6.602^{+ 0.129}_{ 0.000}$ & $-1.060^{+ 0.907}_{-0.134}$ \\
  B017-G070   & $ 9.920^{+ 0.217}_{-0.179}$ & $-2.139^{+ 0.248}_{-0.162}$ \\
  B339-G077   & $10.137^{+ 0.000}_{-0.191}$ & $-1.599^{+ 0.175}_{-0.198}$ \\
  B361-G255   & $ 9.796^{+ 0.199}_{-0.153}$ & $-2.139^{+ 0.261}_{-0.162}$ \\
       G260   & $ 9.827^{+ 0.310}_{-0.144}$ & $ 0.182^{+ 0.216}_{-0.322}$ \\
     SK104A   & $10.137^{+ 0.000}_{-0.088}$ & $-1.545^{+ 0.164}_{-0.198}$ \\
  B396-G335   & $ 7.079^{+ 0.780}_{-0.037}$ & $-1.545^{+ 1.464}_{-0.756}$ \\
  G339-BA30   & $ 6.903^{+ 0.815}_{-0.118}$ & $-1.167^{+ 0.118}_{-0.075}$ \\
  B402-G346   & $10.136^{+ 0.001}_{-0.251}$ & $-1.923^{+ 0.204}_{-0.202}$ \\
      B337D   & $ 6.903^{+ 0.178}_{-0.071}$ & $-1.006^{+ 0.121}_{-0.114}$ \\
        H22   & $10.136^{+ 0.001}_{-0.315}$ & $-0.304^{+ 0.291}_{-0.258}$ \\
 \enddata
\end{deluxetable}

\pagestyle{empty}
\begin{deluxetable}{lrr|rr|rr}
  \tablecolumns{7} \tablewidth{0pc} \tablecaption{Lick/IDS Index
    $\chi^2_{\rm min}$ Fit Results Using \citet{bc03} Models with a
    Chabrier IMF and Padova 1994 Stellar Evolutionary
    Tracks.\label{t7.tab}} \tablehead{\colhead{ID} & \colhead{log $\rm
      Age_{Lick}$} & \colhead{$\rm [Fe/H]_{Lick}$}& \colhead{log $\rm
      Age_{SED}$} & \colhead{$\rm [Fe/H]_{SED}$} & \colhead{log $\rm
      Age_{Both}$} & \colhead{$\rm [Fe/H]_{Both}$} \\ & \colhead{[yr]}
    & \colhead{(dex)} & \colhead{[yr]} & \colhead{(dex)} &
    \colhead{[yr]} & \colhead{(dex)}} \startdata
           H9     & $ 9.916^{+ 0.222}_{-0.220}$ & $-1.743^{+ 0.262}_{-0.298}$ & 	 $ 9.107^{+ 0.479}_{-0.011}$ & $-0.845^{+ 0.226}_{-0.195}$ & 	 $10.000^{+ 0.138}_{-0.375}$ & $-1.687^{+ 0.307}_{-0.494}$ \\          
  MCGC5-H10     & $ 9.845^{+ 0.293}_{-0.172}$ & $-1.743^{+ 0.202}_{-0.334}$ & 	 $ 9.875^{+ 0.263}_{-0.469}$ & $-1.350^{+ 0.305}_{-0.488}$ & 	 $10.031^{+ 0.107}_{-0.383}$ & $-1.800^{+ 0.317}_{-0.449}$ \\          
     SK001A     & $10.097^{+ 0.041}_{-0.176}$ & $-0.676^{+ 0.155}_{-0.209}$ & 	 $ 9.342^{+ 0.473}_{-0.107}$ & $-1.126^{+ 0.373}_{-0.434}$ & 	 $10.097^{+ 0.041}_{-0.306}$ & $-0.901^{+ 0.234}_{-0.281}$ \\          
       B423     & $ 6.440^{+ 0.430}_{-0.277}$ & $ 0.279^{+ 0.280}_{-0.378}$ & 	 $ 8.957^{+ 0.104}_{-0.093}$ & $ 0.559^{+ 0.000}_{-0.066}$ & 	 $10.061^{+ 0.077}_{-0.319}$ & $-1.631^{+ 0.257}_{-0.548}$ \\          
 B298-MCGC6     & $ 9.954^{+ 0.184}_{-0.391}$ & $-2.193^{+ 0.298}_{-0.056}$ & 	 $ 9.796^{+ 0.342}_{-0.613}$ & $-2.193^{+ 0.585}_{-0.056}$ & 	 $ 9.602^{+ 0.536}_{-0.324}$ & $-1.912^{+ 0.353}_{-0.337}$ \\          
        H12     & $ 6.600^{+ 0.093}_{-0.042}$ & $-1.968^{+ 0.321}_{-0.281}$ & 	 $ 9.860^{+ 0.278}_{-0.400}$ & $-2.249^{+ 0.753}_{ 0.000}$ & 	 $ 9.903^{+ 0.235}_{-0.321}$ & $-2.024^{+ 0.373}_{-0.225}$ \\          
      B167D     & $ 6.680^{+ 0.033}_{-0.101}$ & $-2.249^{+ 0.294}_{ 0.000}$ & 	 $ 8.957^{+ 0.087}_{-0.219}$ & $-0.845^{+ 0.303}_{-0.311}$ & 	 $ 6.620^{+ 0.044}_{-0.043}$ & $-2.249^{+ 0.089}_{ 0.000}$ \\          
  B309-G031     & $10.000^{+ 0.138}_{-0.236}$ & $-1.687^{+ 0.196}_{-0.313}$ & 	 $ 9.845^{+ 0.234}_{-0.345}$ & $-1.968^{+ 0.432}_{-0.281}$ & 	 $ 9.966^{+ 0.172}_{-0.283}$ & $-1.743^{+ 0.279}_{-0.394}$ \\          
       B436     & $ 9.107^{+ 0.248}_{-0.057}$ & $-0.283^{+ 0.514}_{-0.248}$ & 	 $ 9.966^{+ 0.172}_{-0.235}$ & $-0.508^{+ 0.236}_{-0.255}$ & 	 $ 9.107^{+ 0.563}_{-0.023}$ & $-0.283^{+ 0.720}_{-0.276}$ \\          
        H15     & $ 9.107^{+ 0.412}_{-0.268}$ & $-1.800^{+ 0.254}_{-0.449}$ & 	 $ 8.907^{+ 0.119}_{-0.064}$ & $ 0.559^{+ 0.000}_{-0.182}$ & 	 $ 9.439^{+ 0.257}_{-0.357}$ & $-1.912^{+ 0.394}_{-0.337}$ \\          
  B006-G058     & $ 9.916^{+ 0.165}_{-0.225}$ & $-0.339^{+ 0.138}_{-0.151}$ & 	 $10.011^{+ 0.127}_{-0.294}$ & $-0.452^{+ 0.215}_{-0.223}$ & 	 $ 9.916^{+ 0.207}_{-0.275}$ & $-0.339^{+ 0.179}_{-0.215}$ \\          
        H16     & $ 6.540^{+ 0.056}_{-0.042}$ & $-2.249^{+ 0.138}_{ 0.000}$ & 	 $ 9.207^{+ 0.131}_{-0.126}$ & $-0.732^{+ 0.290}_{-0.413}$ & 	 $ 9.989^{+ 0.149}_{-0.428}$ & $-1.519^{+ 0.313}_{-0.567}$ \\          
  B017-G070     & $ 9.978^{+ 0.140}_{-0.189}$ & $-1.294^{+ 0.223}_{-0.249}$ & 	 $ 9.889^{+ 0.249}_{-0.214}$ & $-0.452^{+ 0.256}_{-0.243}$ & 	 $ 9.574^{+ 0.276}_{-0.231}$ & $-0.508^{+ 0.223}_{-0.249}$ \\          
  B339-G077     & $ 9.942^{+ 0.159}_{-0.248}$ & $-0.395^{+ 0.151}_{-0.166}$ & 	 $ 9.602^{+ 0.193}_{-0.300}$ & $-0.283^{+ 0.230}_{-0.290}$ & 	 $ 9.845^{+ 0.267}_{-0.380}$ & $-0.395^{+ 0.223}_{-0.238}$ \\          
  B361-G255     & $ 6.820^{+ 0.044}_{-0.027}$ & $-0.283^{+ 0.163}_{-0.183}$ & 	 $ 9.107^{+ 0.478}_{-0.017}$ & $-1.069^{+ 0.212}_{-0.198}$ & 	 $ 9.916^{+ 0.222}_{-0.309}$ & $-1.631^{+ 0.306}_{-0.439}$ \\          
       G260     & $ 7.000^{+ 0.387}_{-0.119}$ & $ 0.279^{+ 0.123}_{-0.450}$ & 	 $ 9.057^{+ 0.031}_{-0.096}$ & $-1.294^{+ 0.338}_{-0.599}$ & 	 $ 9.829^{+ 0.309}_{-0.304}$ & $-1.856^{+ 0.433}_{-0.393}$ \\          
     SK104A     & $ 9.978^{+ 0.160}_{-0.171}$ & $-0.508^{+ 0.127}_{-0.261}$ & 	 $ 9.966^{+ 0.172}_{-0.254}$ & $-0.452^{+ 0.222}_{-0.224}$ & 	 $ 9.978^{+ 0.160}_{-0.228}$ & $-0.508^{+ 0.181}_{-0.265}$ \\          
  B396-G335     & $ 7.180^{+ 0.167}_{-0.156}$ & $-1.182^{+ 0.502}_{-0.339}$ & 	 $ 9.954^{+ 0.184}_{-0.455}$ & $-2.249^{+ 0.717}_{ 0.000}$ & 	 $ 9.966^{+ 0.172}_{-0.378}$ & $-2.080^{+ 0.397}_{-0.169}$ \\          
  G339-BA30     & $ 9.903^{+ 0.235}_{-0.241}$ & $-1.856^{+ 0.287}_{-0.347}$ & 	 $ 9.720^{+ 0.418}_{-0.367}$ & $-1.069^{+ 0.319}_{-0.383}$ & 	 $10.106^{+ 0.032}_{-0.479}$ & $-1.800^{+ 0.354}_{-0.449}$ \\          
  B402-G346     & $ 9.677^{+ 0.217}_{-0.293}$ & $-0.452^{+ 0.201}_{-0.169}$ & 	 $ 9.107^{+ 0.626}_{-0.007}$ & $-0.564^{+ 0.852}_{-0.228}$ & 	 $ 9.628^{+ 0.238}_{-0.317}$ & $-0.452^{+ 0.231}_{-0.286}$ \\          
      B337D     & $ 9.966^{+ 0.172}_{-0.247}$ & $-1.575^{+ 0.211}_{-0.292}$ & 	 $ 8.957^{+ 0.085}_{-0.105}$ & $ 0.559^{+ 0.000}_{-0.050}$ & 	 $10.041^{+ 0.097}_{-0.394}$ & $-1.406^{+ 0.264}_{-0.362}$ \\          
        H22     & $10.088^{+ 0.050}_{-0.450}$ & $-2.024^{+ 0.357}_{-0.225}$ & 	 $ 9.301^{+ 0.450}_{-0.202}$ & $-0.901^{+ 0.339}_{-0.435}$ & 	 $10.130^{+ 0.008}_{-0.544}$ & $-1.912^{+ 0.378}_{-0.337}$ \\          
 \enddata
\end{deluxetable}

\addtocounter{table}{-1}
\begin{deluxetable}{lrr|rr|rr}
  \tablecolumns{7} \tablewidth{0pc} \tablecaption{Continued, but for a
    Salpeter IMF and Padova 1994 Stellar Evolutionary Tracks.}
  \tablehead{\colhead{ID} & \colhead{log $\rm Age_{Lick}$} &
    \colhead{$\rm [Fe/H]_{Lick}$}& \colhead{log $\rm Age_{SED}$} &
    \colhead{$\rm [Fe/H]_{SED}$} & \colhead{log $\rm Age_{Both}$} &
    \colhead{$\rm [Fe/H]_{Both}$} \\ & \colhead{[yr]} &
    \colhead{(dex)} & \colhead{[yr]} & \colhead{(dex)} &
    \colhead{[yr]} & \colhead{(dex)}} \startdata
           H9     & $ 6.600^{+ 0.100}_{-0.053}$ & $-2.249^{+ 0.291}_{ 0.000}$ & 	 $ 9.107^{+ 0.451}_{-0.011}$ & $-0.845^{+ 0.226}_{-0.202}$ & 	 $10.000^{+ 0.138}_{-0.376}$ & $-1.743^{+ 0.313}_{-0.506}$ \\          
  MCGC5-H10     & $ 9.929^{+ 0.209}_{-0.238}$ & $-1.856^{+ 0.246}_{-0.289}$ & 	 $10.138^{+ 0.000}_{-0.532}$ & $-1.687^{+ 0.468}_{-0.562}$ & 	 $ 9.929^{+ 0.209}_{-0.308}$ & $-1.800^{+ 0.298}_{-0.449}$ \\          
     SK001A     & $10.097^{+ 0.041}_{-0.232}$ & $-0.676^{+ 0.166}_{-0.237}$ & 	 $ 9.342^{+ 0.294}_{-0.114}$ & $-1.126^{+ 0.351}_{-0.478}$ & 	 $10.000^{+ 0.138}_{-0.219}$ & $-0.957^{+ 0.278}_{-0.317}$ \\          
       B423     & $ 6.440^{+ 0.434}_{-0.312}$ & $ 0.279^{+ 0.280}_{-0.377}$ & 	 $ 8.957^{+ 0.103}_{-0.092}$ & $ 0.559^{+ 0.000}_{-0.067}$ & 	 $10.088^{+ 0.050}_{-0.315}$ & $-1.743^{+ 0.312}_{-0.506}$ \\          
 B298-MCGC6     & $ 9.544^{+ 0.219}_{-0.167}$ & $-1.856^{+ 0.277}_{-0.348}$ & 	 $ 9.255^{+ 0.414}_{-0.146}$ & $-1.631^{+ 0.582}_{-0.618}$ & 	 $ 9.544^{+ 0.369}_{-0.293}$ & $-1.856^{+ 0.347}_{-0.393}$ \\          
        H12     & $ 6.600^{+ 0.102}_{-0.060}$ & $-2.024^{+ 0.373}_{-0.225}$ & 	 $ 9.813^{+ 0.325}_{-0.462}$ & $-2.249^{+ 0.735}_{ 0.000}$ & 	 $ 9.813^{+ 0.325}_{-0.270}$ & $-2.024^{+ 0.387}_{-0.225}$ \\          
      B167D     & $ 6.680^{+ 0.032}_{-0.103}$ & $-2.249^{+ 0.289}_{ 0.000}$ & 	 $ 8.957^{+ 0.084}_{-0.225}$ & $-0.845^{+ 0.295}_{-0.333}$ & 	 $ 6.600^{+ 3.494}_{-0.022}$ & $-2.249^{+ 0.083}_{ 0.000}$ \\          
  B309-G031     & $10.000^{+ 0.138}_{-0.240}$ & $-1.743^{+ 0.215}_{-0.306}$ & 	 $ 9.829^{+ 0.203}_{-0.430}$ & $-1.968^{+ 0.418}_{-0.281}$ & 	 $ 9.916^{+ 0.222}_{-0.238}$ & $-1.800^{+ 0.317}_{-0.386}$ \\          
       B436     & $ 9.107^{+ 0.246}_{-0.059}$ & $-0.283^{+ 0.544}_{-0.250}$ & 	 $ 9.966^{+ 0.172}_{-0.219}$ & $-0.564^{+ 0.251}_{-0.257}$ & 	 $ 9.107^{+ 0.539}_{-0.024}$ & $-0.283^{+ 0.724}_{-0.277}$ \\          
        H15     & $ 9.107^{+ 0.388}_{-0.278}$ & $-1.800^{+ 0.250}_{-0.449}$ & 	 $ 8.907^{+ 0.116}_{-0.064}$ & $ 0.559^{+ 0.000}_{-0.188}$ & 	 $ 9.415^{+ 0.232}_{-0.353}$ & $-1.856^{+ 0.343}_{-0.393}$ \\          
  B006-G058     & $ 9.903^{+ 0.168}_{-0.231}$ & $-0.339^{+ 0.145}_{-0.146}$ & 	 $ 9.966^{+ 0.172}_{-0.257}$ & $-0.452^{+ 0.222}_{-0.227}$ & 	 $ 9.903^{+ 0.198}_{-0.315}$ & $-0.339^{+ 0.180}_{-0.213}$ \\          
        H16     & $ 6.540^{+ 0.056}_{-0.042}$ & $-2.249^{+ 0.138}_{ 0.000}$ & 	 $ 9.207^{+ 0.104}_{-0.132}$ & $-0.732^{+ 0.274}_{-0.466}$ & 	 $ 9.966^{+ 0.172}_{-0.374}$ & $-1.631^{+ 0.396}_{-0.538}$ \\          
  B017-G070     & $ 9.989^{+ 0.124}_{-0.170}$ & $-1.406^{+ 0.260}_{-0.189}$ & 	 $ 9.829^{+ 0.309}_{-0.217}$ & $-0.452^{+ 0.257}_{-0.215}$ & 	 $ 9.574^{+ 0.249}_{-0.245}$ & $-0.508^{+ 0.225}_{-0.265}$ \\          
  B339-G077     & $ 9.875^{+ 0.185}_{-0.244}$ & $-0.339^{+ 0.136}_{-0.199}$ & 	 $ 9.574^{+ 0.285}_{-0.254}$ & $-0.339^{+ 0.257}_{-0.253}$ & 	 $ 9.760^{+ 0.324}_{-0.321}$ & $-0.395^{+ 0.242}_{-0.196}$ \\          
  B361-G255     & $ 6.820^{+ 0.043}_{-0.027}$ & $-0.283^{+ 0.161}_{-0.182}$ & 	 $ 9.107^{+ 0.458}_{-0.018}$ & $-1.069^{+ 0.209}_{-0.210}$ & 	 $ 9.929^{+ 0.209}_{-0.321}$ & $-1.687^{+ 0.313}_{-0.473}$ \\          
       G260     & $ 6.980^{+ 0.400}_{-0.099}$ & $ 0.279^{+ 0.143}_{-0.452}$ & 	 $ 9.057^{+ 0.028}_{-0.116}$ & $-1.294^{+ 0.324}_{-0.732}$ & 	 $ 9.829^{+ 0.309}_{-0.308}$ & $-1.912^{+ 0.453}_{-0.337}$ \\          
     SK104A     & $ 9.966^{+ 0.152}_{-0.196}$ & $-0.508^{+ 0.142}_{-0.237}$ & 	 $ 9.829^{+ 0.309}_{-0.179}$ & $-0.395^{+ 0.254}_{-0.225}$ & 	 $ 9.966^{+ 0.172}_{-0.246}$ & $-0.508^{+ 0.190}_{-0.264}$ \\          
  B396-G335     & $ 7.180^{+ 0.164}_{-0.268}$ & $-1.182^{+ 0.511}_{-0.337}$ & 	 $ 9.796^{+ 0.342}_{-0.346}$ & $-2.249^{+ 0.801}_{ 0.000}$ & 	 $ 9.889^{+ 0.249}_{-0.341}$ & $-2.080^{+ 0.397}_{-0.169}$ \\          
  G339-BA30     & $ 9.829^{+ 0.309}_{-0.189}$ & $-1.856^{+ 0.291}_{-0.393}$ & 	 $ 9.699^{+ 0.372}_{-0.318}$ & $-1.126^{+ 0.325}_{-0.404}$ & 	 $10.079^{+ 0.059}_{-0.454}$ & $-1.856^{+ 0.355}_{-0.393}$ \\          
  B402-G346     & $ 9.628^{+ 0.178}_{-0.268}$ & $-0.395^{+ 0.167}_{-0.206}$ & 	 $ 9.107^{+ 0.597}_{-0.007}$ & $-0.564^{+ 0.846}_{-0.230}$ & 	 $ 9.628^{+ 0.184}_{-0.350}$ & $-0.452^{+ 0.226}_{-0.301}$ \\          
      B337D     & $10.000^{+ 0.138}_{-0.272}$ & $-1.631^{+ 0.214}_{-0.291}$ & 	 $ 8.957^{+ 0.085}_{-0.104}$ & $ 0.559^{+ 0.000}_{-0.051}$ & 	 $10.061^{+ 0.077}_{-0.347}$ & $-1.519^{+ 0.303}_{-0.365}$ \\          
        H22     & $ 9.813^{+ 0.325}_{-0.197}$ & $-2.024^{+ 0.352}_{-0.225}$ & 	 $ 9.301^{+ 0.297}_{-0.203}$ & $-0.901^{+ 0.322}_{-0.471}$ & 	 $10.138^{+ 0.000}_{-0.517}$ & $-2.080^{+ 0.482}_{-0.169}$ \\          
 \enddata
\end{deluxetable}

\addtocounter{table}{-1}
\begin{deluxetable}{lrr|rr|rr}
  \tablecolumns{7} \tablewidth{0pc} \tablecaption{Continued, but for a
    Chabrier IMF and Padova 2000 Stellar Evolutionary Tracks.}
  \tablehead{\colhead{ID} & \colhead{log $\rm Age_{Lick}$} &
    \colhead{$\rm [Fe/H]_{Lick}$}& \colhead{log $\rm Age_{SED}$} &
    \colhead{$\rm [Fe/H]_{SED}$} & \colhead{log $\rm Age_{Both}$} &
    \colhead{$\rm [Fe/H]_{Both}$} \\ & \colhead{[yr]} &
    \colhead{(dex)} & \colhead{[yr]} & \colhead{(dex)} &
    \colhead{[yr]} & \colhead{(dex)}} \startdata
           H9     & $ 9.954^{+ 0.130}_{-0.251}$ & $-1.647^{+ 0.076}_{ 0.000}$ & 	 $ 9.157^{+ 0.394}_{-0.009}$ & $-0.757^{+ 0.139}_{-0.156}$ & 	 $ 9.954^{+ 0.184}_{-0.301}$ & $-1.647^{+ 0.126}_{ 0.000}$ \\          
  MCGC5-H10     & $ 9.875^{+ 0.198}_{-0.188}$ & $-1.647^{+ 0.056}_{ 0.000}$ & 	 $10.070^{+ 0.068}_{-0.489}$ & $-1.492^{+ 0.432}_{-0.155}$ & 	 $ 9.875^{+ 0.263}_{-0.244}$ & $-1.647^{+ 0.101}_{ 0.000}$ \\          
     SK001A     & $10.070^{+ 0.068}_{-0.200}$ & $-0.602^{+ 0.109}_{-0.172}$ & 	 $ 9.829^{+ 0.309}_{-0.512}$ & $-1.337^{+ 0.298}_{-0.310}$ & 	 $ 9.989^{+ 0.149}_{-0.236}$ & $-0.834^{+ 0.283}_{-0.423}$ \\          
       B423     & $ 6.500^{+ 0.065}_{-0.183}$ & $ 0.211^{+ 0.077}_{-0.420}$ & 	 $ 9.157^{+ 0.010}_{-0.004}$ & $-0.370^{+ 0.156}_{-0.144}$ & 	 $10.011^{+ 0.127}_{-0.287}$ & $-1.647^{+ 0.083}_{ 0.000}$ \\          
 B298-MCGC6     & $ 9.477^{+ 0.313}_{-0.187}$ & $-1.647^{+ 0.050}_{ 0.000}$ & 	 $ 9.342^{+ 0.522}_{-0.182}$ & $-1.647^{+ 0.494}_{ 0.000}$ & 	 $ 9.439^{+ 0.413}_{-0.206}$ & $-1.647^{+ 0.097}_{ 0.000}$ \\          
        H12     & $ 6.620^{+ 0.054}_{-0.031}$ & $-1.608^{+ 0.104}_{-0.039}$ & 	 $ 9.699^{+ 0.391}_{-0.463}$ & $-1.647^{+ 0.291}_{ 0.000}$ & 	 $ 9.720^{+ 0.414}_{-0.297}$ & $-1.647^{+ 0.065}_{ 0.000}$ \\          
      B167D     & $ 6.620^{+ 0.037}_{-0.021}$ & $-1.608^{+ 0.155}_{-0.039}$ & 	 $ 9.007^{+ 0.086}_{-0.214}$ & $-0.757^{+ 0.283}_{-0.213}$ & 	 $ 9.699^{+ 0.375}_{-0.263}$ & $-1.647^{+ 0.089}_{ 0.000}$ \\          
  B309-G031     & $ 9.966^{+ 0.128}_{-0.209}$ & $-1.647^{+ 0.062}_{ 0.000}$ & 	 $ 9.813^{+ 0.215}_{-0.531}$ & $-1.647^{+ 0.251}_{ 0.000}$ & 	 $ 9.966^{+ 0.165}_{-0.292}$ & $-1.647^{+ 0.099}_{ 0.000}$ \\          
       B436     & $ 9.107^{+ 0.218}_{-0.042}$ & $-0.137^{+ 0.247}_{-0.287}$ & 	 $10.138^{+ 0.000}_{-0.323}$ & $-0.524^{+ 0.174}_{-0.224}$ & 	 $ 9.157^{+ 0.274}_{-0.038}$ & $-0.215^{+ 0.330}_{-0.333}$ \\          
        H15     & $ 9.057^{+ 0.390}_{-0.228}$ & $-1.647^{+ 0.194}_{ 0.000}$ & 	 $ 9.107^{+ 0.149}_{-0.047}$ & $ 0.017^{+ 0.271}_{-0.303}$ & 	 $ 9.342^{+ 0.258}_{-0.294}$ & $-1.647^{+ 0.097}_{ 0.000}$ \\          
  B006-G058     & $ 9.889^{+ 0.249}_{-0.256}$ & $-0.370^{+ 0.162}_{-0.156}$ & 	 $ 9.929^{+ 0.209}_{-0.346}$ & $-0.292^{+ 0.238}_{-0.228}$ & 	 $ 9.889^{+ 0.249}_{-0.296}$ & $-0.370^{+ 0.245}_{-0.181}$ \\          
        H16     & $ 6.620^{+ 0.024}_{-0.029}$ & $-1.182^{+ 0.265}_{-0.302}$ & 	 $ 9.230^{+ 0.136}_{-0.059}$ & $-0.641^{+ 0.316}_{-0.526}$ & 	 $ 9.954^{+ 0.184}_{-0.360}$ & $-1.608^{+ 0.221}_{-0.039}$ \\          
  B017-G070     & $ 9.966^{+ 0.085}_{-0.216}$ & $-1.415^{+ 0.165}_{-0.133}$ & 	 $ 9.796^{+ 0.316}_{-0.309}$ & $-0.292^{+ 0.255}_{-0.233}$ & 	 $ 9.628^{+ 0.260}_{-0.252}$ & $-0.563^{+ 0.176}_{-0.250}$ \\          
  B339-G077     & $ 9.860^{+ 0.278}_{-0.222}$ & $-0.408^{+ 0.177}_{-0.158}$ & 	 $ 9.230^{+ 0.160}_{-0.085}$ & $ 0.211^{+ 0.077}_{-0.213}$ & 	 $ 9.860^{+ 0.278}_{-0.322}$ & $-0.408^{+ 0.221}_{-0.218}$ \\          
  B361-G255     & $ 6.820^{+ 0.044}_{-0.027}$ & $-0.292^{+ 0.156}_{-0.170}$ & 	 $ 9.225^{+ 0.171}_{-0.053}$ & $-0.757^{+ 0.278}_{-0.377}$ & 	 $ 9.954^{+ 0.184}_{-0.363}$ & $-1.608^{+ 0.166}_{-0.039}$ \\          
       G260     & $ 7.200^{+ 0.250}_{-0.251}$ & $ 0.017^{+ 0.242}_{-0.178}$ & 	 $ 9.107^{+ 0.041}_{-0.063}$ & $-1.144^{+ 0.358}_{-0.503}$ & 	 $ 9.544^{+ 0.361}_{-0.281}$ & $-1.531^{+ 0.170}_{-0.116}$ \\          
     SK104A     & $ 9.989^{+ 0.149}_{-0.183}$ & $-0.563^{+ 0.124}_{-0.178}$ & 	 $10.122^{+ 0.016}_{-0.357}$ & $-0.447^{+ 0.192}_{-0.210}$ & 	 $10.051^{+ 0.087}_{-0.284}$ & $-0.524^{+ 0.145}_{-0.205}$ \\          
  B396-G335     & $ 7.180^{+ 0.218}_{-0.326}$ & $-1.066^{+ 0.438}_{-0.499}$ & 	 $ 9.677^{+ 0.461}_{-0.433}$ & $-1.647^{+ 0.387}_{ 0.000}$ & 	 $ 9.720^{+ 0.418}_{-0.350}$ & $-1.647^{+ 0.064}_{ 0.000}$ \\          
  G339-BA30     & $ 9.813^{+ 0.325}_{-0.210}$ & $-1.647^{+ 0.076}_{ 0.000}$ & 	 $ 9.398^{+ 0.295}_{-0.174}$ & $-0.718^{+ 0.255}_{-0.367}$ & 	 $10.021^{+ 0.117}_{-0.437}$ & $-1.647^{+ 0.104}_{ 0.000}$ \\          
  B402-G346     & $ 9.699^{+ 0.236}_{-0.304}$ & $-0.486^{+ 0.150}_{-0.199}$ & 	 $ 9.157^{+ 0.498}_{-0.006}$ & $-0.563^{+ 0.591}_{-0.168}$ & 	 $ 9.512^{+ 0.395}_{-0.189}$ & $-0.447^{+ 0.286}_{-0.250}$ \\          
      B337D     & $ 9.954^{+ 0.129}_{-0.212}$ & $-1.608^{+ 0.142}_{-0.039}$ & 	 $ 9.157^{+ 0.008}_{-0.004}$ & $-0.370^{+ 0.124}_{-0.122}$ & 	 $ 9.157^{+ 0.275}_{-0.013}$ & $-0.524^{+ 0.303}_{-0.200}$ \\          
        H22     & $ 9.699^{+ 0.439}_{-0.215}$ & $-1.647^{+ 0.056}_{ 0.000}$ & 	 $ 9.574^{+ 0.498}_{-0.294}$ & $-1.066^{+ 0.345}_{-0.450}$ & 	 $10.130^{+ 0.008}_{-0.680}$ & $-1.647^{+ 0.069}_{ 0.000}$ \\          
 \enddata
\end{deluxetable}

\addtocounter{table}{-1}
\begin{deluxetable}{lrr|rr|rr}
  \tablecolumns{7} \tablewidth{0pc} \tablecaption{Continued, but for a
    Salpeter IMF and Padova 2000 Stellar Evolutionary Tracks.}
  \tablehead{\colhead{ID} & \colhead{log $\rm Age_{Lick}$} &
    \colhead{$\rm [Fe/H]_{Lick}$}& \colhead{log $\rm Age_{SED}$} &
    \colhead{$\rm [Fe/H]_{SED}$} & \colhead{log $\rm Age_{Both}$} &
    \colhead{$\rm [Fe/H]_{Both}$} \\ & \colhead{[yr]} &
    \colhead{(dex)} & \colhead{[yr]} & \colhead{(dex)} &
    \colhead{[yr]} & \colhead{(dex)}} \startdata
           H9     & $ 9.875^{+ 0.197}_{-0.188}$ & $-1.647^{+ 0.069}_{ 0.000}$ & 	 $ 9.207^{+ 0.153}_{-0.016}$ & $-0.486^{+ 0.334}_{-0.208}$ & 	 $ 9.942^{+ 0.196}_{-0.319}$ & $-1.647^{+ 0.106}_{ 0.000}$ \\          
  MCGC5-H10     & $ 9.829^{+ 0.216}_{-0.157}$ & $-1.647^{+ 0.052}_{ 0.000}$ & 	 $10.079^{+ 0.059}_{-0.452}$ & $-1.647^{+ 0.480}_{ 0.000}$ & 	 $ 9.875^{+ 0.263}_{-0.273}$ & $-1.647^{+ 0.087}_{ 0.000}$ \\          
     SK001A     & $10.051^{+ 0.087}_{-0.206}$ & $-0.602^{+ 0.117}_{-0.175}$ & 	 $ 9.602^{+ 0.501}_{-0.319}$ & $-1.299^{+ 0.362}_{-0.348}$ & 	 $ 9.989^{+ 0.149}_{-0.291}$ & $-0.834^{+ 0.274}_{-0.492}$ \\          
       B423     & $ 6.500^{+ 0.326}_{-0.189}$ & $ 0.172^{+ 0.116}_{-0.375}$ & 	 $ 9.157^{+ 0.010}_{-0.005}$ & $-0.370^{+ 0.162}_{-0.144}$ & 	 $10.021^{+ 0.117}_{-0.313}$ & $-1.647^{+ 0.069}_{ 0.000}$ \\          
 B298-MCGC6     & $ 9.439^{+ 0.285}_{-0.162}$ & $-1.647^{+ 0.051}_{ 0.000}$ & 	 $ 9.279^{+ 0.426}_{-0.129}$ & $-1.569^{+ 0.539}_{-0.078}$ & 	 $ 9.439^{+ 0.364}_{-0.216}$ & $-1.647^{+ 0.090}_{ 0.000}$ \\          
        H12     & $ 6.620^{+ 0.053}_{-0.031}$ & $-1.608^{+ 0.110}_{-0.039}$ & 	 $ 9.677^{+ 0.311}_{-0.458}$ & $-1.647^{+ 0.267}_{ 0.000}$ & 	 $ 9.699^{+ 0.336}_{-0.334}$ & $-1.647^{+ 0.060}_{ 0.000}$ \\          
      B167D     & $ 6.620^{+ 0.044}_{-0.023}$ & $-1.569^{+ 0.124}_{-0.078}$ & 	 $ 9.007^{+ 0.089}_{-0.198}$ & $-0.795^{+ 0.324}_{-0.189}$ & 	 $ 9.699^{+ 0.290}_{-0.312}$ & $-1.647^{+ 0.080}_{ 0.000}$ \\          
  B309-G031     & $ 9.954^{+ 0.135}_{-0.216}$ & $-1.647^{+ 0.052}_{ 0.000}$ & 	 $ 9.813^{+ 0.176}_{-0.597}$ & $-1.647^{+ 0.241}_{ 0.000}$ & 	 $ 9.875^{+ 0.246}_{-0.224}$ & $-1.647^{+ 0.098}_{ 0.000}$ \\          
       B436     & $ 9.107^{+ 0.211}_{-0.042}$ & $-0.137^{+ 0.246}_{-0.284}$ & 	 $10.138^{+ 0.000}_{-0.296}$ & $-0.602^{+ 0.207}_{-0.231}$ & 	 $ 9.157^{+ 0.246}_{-0.042}$ & $-0.176^{+ 0.291}_{-0.361}$ \\          
        H15     & $ 9.057^{+ 0.343}_{-0.230}$ & $-1.647^{+ 0.190}_{ 0.000}$ & 	 $ 9.107^{+ 0.114}_{-0.060}$ & $ 0.056^{+ 0.232}_{-0.352}$ & 	 $ 9.301^{+ 0.289}_{-0.263}$ & $-1.647^{+ 0.106}_{ 0.000}$ \\          
  B006-G058     & $ 9.845^{+ 0.229}_{-0.235}$ & $-0.370^{+ 0.186}_{-0.144}$ & 	 $ 9.829^{+ 0.273}_{-0.320}$ & $-0.254^{+ 0.237}_{-0.233}$ & 	 $ 9.845^{+ 0.293}_{-0.313}$ & $-0.331^{+ 0.223}_{-0.210}$ \\          
        H16     & $ 6.620^{+ 0.025}_{-0.030}$ & $-1.182^{+ 0.265}_{-0.301}$ & 	 $ 9.230^{+ 0.135}_{-0.070}$ & $-0.679^{+ 0.329}_{-0.550}$ & 	 $ 9.942^{+ 0.196}_{-0.332}$ & $-1.647^{+ 0.225}_{ 0.000}$ \\          
  B017-G070     & $ 9.954^{+ 0.098}_{-0.195}$ & $-1.453^{+ 0.158}_{-0.122}$ & 	 $ 9.796^{+ 0.310}_{-0.296}$ & $-0.331^{+ 0.254}_{-0.220}$ & 	 $ 9.628^{+ 0.235}_{-0.258}$ & $-0.563^{+ 0.172}_{-0.268}$ \\          
  B339-G077     & $ 9.813^{+ 0.244}_{-0.238}$ & $-0.370^{+ 0.164}_{-0.177}$ & 	 $ 9.255^{+ 0.150}_{-0.100}$ & $ 0.172^{+ 0.116}_{-0.245}$ & 	 $ 9.813^{+ 0.268}_{-0.327}$ & $-0.370^{+ 0.201}_{-0.241}$ \\          
  B361-G255     & $ 6.820^{+ 0.040}_{-0.026}$ & $-0.292^{+ 0.148}_{-0.167}$ & 	 $ 9.207^{+ 0.148}_{-0.024}$ & $-0.718^{+ 0.219}_{-0.352}$ & 	 $ 9.875^{+ 0.263}_{-0.310}$ & $-1.608^{+ 0.162}_{-0.039}$ \\          
       G260     & $ 7.200^{+ 0.243}_{-0.257}$ & $ 0.017^{+ 0.240}_{-0.171}$ & 	 $ 9.107^{+ 0.055}_{-0.063}$ & $-1.182^{+ 0.383}_{-0.465}$ & 	 $ 9.544^{+ 0.356}_{-0.266}$ & $-1.569^{+ 0.189}_{-0.078}$ \\          
     SK104A     & $ 9.978^{+ 0.160}_{-0.204}$ & $-0.563^{+ 0.136}_{-0.169}$ & 	 $10.130^{+ 0.008}_{-0.353}$ & $-0.486^{+ 0.186}_{-0.215}$ & 	 $10.021^{+ 0.117}_{-0.295}$ & $-0.524^{+ 0.153}_{-0.217}$ \\          
  B396-G335     & $ 7.180^{+ 0.215}_{-0.329}$ & $-1.066^{+ 0.437}_{-0.491}$ & 	 $ 9.544^{+ 0.495}_{-0.314}$ & $-1.647^{+ 0.499}_{ 0.000}$ & 	 $ 9.699^{+ 0.363}_{-0.358}$ & $-1.647^{+ 0.060}_{ 0.000}$ \\          
  G339-BA30     & $ 7.300^{+ 0.145}_{-0.333}$ & $-0.912^{+ 0.329}_{-0.222}$ & 	 $ 9.380^{+ 0.280}_{-0.163}$ & $-0.718^{+ 0.265}_{-0.373}$ & 	 $ 9.875^{+ 0.263}_{-0.317}$ & $-1.647^{+ 0.103}_{ 0.000}$ \\          
  B402-G346     & $ 9.653^{+ 0.244}_{-0.267}$ & $-0.486^{+ 0.160}_{-0.187}$ & 	 $ 9.157^{+ 0.450}_{-0.007}$ & $-0.563^{+ 0.600}_{-0.170}$ & 	 $ 9.544^{+ 0.321}_{-0.228}$ & $-0.447^{+ 0.238}_{-0.278}$ \\          
      B337D     & $ 9.954^{+ 0.126}_{-0.223}$ & $-1.608^{+ 0.109}_{-0.039}$ & 	 $ 9.157^{+ 0.008}_{-0.004}$ & $-0.370^{+ 0.129}_{-0.121}$ & 	 $ 9.157^{+ 0.261}_{-0.014}$ & $-0.524^{+ 0.328}_{-0.201}$ \\          
        H22     & $ 9.699^{+ 0.216}_{-0.237}$ & $-1.647^{+ 0.048}_{ 0.000}$ & 	 $ 9.544^{+ 0.395}_{-0.275}$ & $-1.066^{+ 0.344}_{-0.472}$ & 	 $ 9.699^{+ 0.439}_{-0.270}$ & $-1.647^{+ 0.090}_{ 0.000}$ \\          
 \enddata
\end{deluxetable}

\clearpage

\label{lastpage}
\end{document}